\documentclass[11pt]{article}
\pdfoutput=1 
\usepackage{jheppub}
\usepackage{graphicx,verbatim, upgreek, txfonts, yfonts,mathabx, amsmath}
\usepackage{psfrag, times}
\usepackage[T1]{fontenc}
\usepackage{extarrows}

\def\a{\alpha}
\def\b{\beta}
\def\D{\Delta}
\def\e{\epsilon}
\def\ka{\kappa}

\def\lam{\lambda}
\def\ga{\gamma}
\def\w{\omega}

\def\nn{\nonumber}
\def\({\left(}
\def\){\right)}
\def\[{\left[}
\def\]{\right]}
\def\pl{\prod\limits}
\def\cN{{\mathcal N}}
\def\cI{{\mathcal I}}

\def\cW{{\mathcal{W}}}
\def\be{\begin{equation}}
\def\ee{\end{equation}}

\def\cQ{{\mathcal{Q}}}
\def\cF{{\mathcal{F}}}
\def\ba{{\bf{a}}}
\def\fa{\mathfrak{a}}
\def\fp{{\mathfrak{p}}}
\def\fq{\mathfrak{q}}
\def\ft{\mathfrak{t}}
\def\fR{\mathfrak{R}}

\def\fr{\mathfrak{r}}
\def\tfr{\tilde{\mathfrak{r}}}
\def\cG{{\mathcal{G}}}
\def\ta{\tilde{a}}
\def\tb{\tilde{b}}
\def\tc{\tilde{c}}
\def\td{\tilde{d}}
\def\tN{\tilde{N}}
\def\tQ{\tilde{Q}}

\def\tu{\tilde{u}}
\def\bbT{{\mathbb T}}
\def\hn{\hat{n}}
\def\tz{\tilde{z}}
\def\tfa{\tilde{\mathfrak a}}
\def\ep{\epsilon}
\def\cM{{\mathcal{M}}}

\def\a{\alpha}
\def\e{\epsilon}

\def\w{\omega}
\def\s{\sigma}

\def\tr{\tilde{r}}
\def\tu{\tilde{u}}
\def\rq{{\mathrm q}}
\def\ta{\tilde{a}}
\def\tb{\tilde{b}}

\def\a{\alpha}
\def\b{\beta}
\def\D{\Delta}
\def\e{\epsilon}
\def\ka{\kappa}

\def\lam{\lambda}
\def\ga{\gamma}
\def\w{\omega}

\def\nn{\nonumber}
\def\({\left(}
\def\){\right)}
\def\[{\left[}
\def\]{\right]}
\def\pl{\prod\limits}
\def\cN{{\mathcal N}}
\def\cI{{\mathcal I}}

\def\cW{{\mathcal{W}}}
\def\be{\begin{equation}}
\def\ee{\end{equation}}

\def\cQ{{\mathcal{Q}}}
\def\ba{{\bf{a}}}
\def\fa{\mathfrak{a}}
\def\fb{\mathfrak{b}}
\def\fp{{\mathfrak{p}}}
\def\fq{\mathfrak{q}}
\def\ft{\mathfrak{t}}
\def\fR{\mathfrak{R}}

\def\fr{\mathfrak{r}}
\def\tfr{\tilde{\mathfrak{r}}}
\def\cG{{\mathcal{G}}}
\def\ta{\tilde{a}}
\def\tb{\tilde{b}}
\def\tc{\tilde{c}}
\def\td{\tilde{d}}
\def\tN{\tilde{N}}
\def\tQ{\tilde{Q}}

\def\tu{\tilde{u}}
\def\bbT{{\mathbb T}}
\def\hn{\hat{n}}
\def\tz{\tilde{z}}
\def\tfa{\tilde{\mathfrak a}}
\def\ep{\epsilon}
\def\bG{\mathbb{G}}

\def\bL{\mathbb{L}}
\def\a{\alpha}
\def\e{\epsilon}

\def\w{\omega}
\def\s{\sigma}

\def\tr{\tilde{r}}
\def\tu{\tilde{u}}
\def\rq{{\mathrm q}}

\begin{document}

\title{\LARGE \bf Heterotic Surface Defects and Dualities from 2d/4d Indices}

\author{ Heng-Yu Chen and Hsiao-Yi Chen}
\affiliation{ Department of Physics and Center for Theoretical Sciences, \\
National Taiwan University, Taipei 10617, Taiwan}
\emailAdd{ heng.yu.chen@phys.ntu.edu.tw}
\emailAdd{chenhsiaoyi1025@gmail.com}
\vspace{2cm}

\abstract{Starting with the superconformal indices for 4d $\cN=2$ and $\cN=1$ supersymmetric gauge theories, which are related by superpotential deformation, 
we perform the contour integrations and isolate the residue contributions which can be attributed to the surface defects. These defects can be interpreted as the IR limit of dynamical vortices.
Given the 2d $\cN=(2,2)$ and $\cN=(0,2)$ world sheet theories for these surface defects,
we then verify this statement by explicitly computing their elliptic genera and identifying their fugacity parameters through superconformal algebras. 
We show them precisely match, and the results extend previous prescriptions for inserting surface defects into 4d supersymmetric partition functions to $\cN=1$ setting. 
We also study how 4d $\cN=1$ IR dualities descend into the $\cN=(0,2)$ world sheet theories of their surface defects, and extend the $\cN=(2,2)$ triality observed earlier
to other $\cN=(0,2)$ surface defects unrelated to dynamical vortices. }

\maketitle

\section{Introduction and Summary}\label{Introduction}
\paragraph{}
Over the past few years, we have witnessed tremendous progress in computing partition functions of supersymmetric gauge theories on compact manifolds in different dimensions using localization techniques.
{These exact results encode different useful information depending on the specific geometry.}
There are many exciting applications such as verifying field theoretic IR dualities e. g. \cite{Seiberg:1994pq, Kutasov:1995np}, the correspondences with 2d CFTs on Riemann surfaces \cite{Alday:2009aq} or Chern-Simons theory on hyperbolic three manifolds \cite{Dimofte:2011ju, Dimofte:2011py}; another interesting class of physical observables which can be studied through these partition functions are co-dimension two defects, which are BPS objects and, in four dimensions, usually called {\it ``surface defects''}.
{There are various possible surface defects, for which, along their two dimensional world sheet ${\mathcal{S}}$, generally part of the bulk 4d gauge group $\bL \subseteq \bG$ is preserved ( $\bL$ is called ``Levi Group''),} and whose presence can be encoded by the singularities in 4d gauge fields.
See for example \cite{Alday:2009fs, Gaiotto:2009fs, Gadde:2013dda, Gaiotto:2013sma} for recent discussions.
As the gauge symmetry breaking occurs in the transverse co-dimensions away from the surface defects,{we can parameterize them using the spatially dependent vacuum expectation values of baryonic operators, which vanish at the positions of the surface operators where the gauge symmetry is partially restored \cite{Gaiotto:2012xa}.}
Perhaps the most well studied surface operators are non-abelian vortices in 4d $\cN=2$ supersymmetric gauge theories (or more precisely their IR limit) considered in \cite{Hanany:2004ea, Shifman:2004dr}, where the vortex world sheet can be readily deduced to be 2d $\cN=(2,2)$ Gauged Linear Sigma Model (GLSM).
Moreover, we can break 4d $\cN=2$ supersymmetry down to $\cN=1$ by introducing appropriate superpotential deformation, while vortices in the resultant theory described by a 2d $\cN=(0,2)$ GLSM are now dubbed {\it ``Heterotic vortices''} \cite{Edalati:2007vk, Tong:2007qj}. In particular, the 4d bulk SUSY breaking pattern can also descend into the 2d vortex world sheet, which we will review in this introduction momentarily.
\paragraph{}
A class of exact 4d partition functions particularly suitable for studying these surface defects are superconformal indices of four dimensional $\cN=1$ and $\cN=2$ supersymmetric field theories \cite{Romelsberger:2005eg, Kinney:2005ej, Gadde:2011ik, Gadde:2011uv} \footnote{Superconformal index can also be defined for $\cN=4$ superconformal field theory which features prominently in the non-trivial checks of AdS/CFT correspondence.}, which trace over the appropriate subset of Hilbert space obtained from radial quantization depending on the choice of truncation condition. 
Although, geometrically due to the background fluxes for the global symmetries, we {can} regard them as the twisted partition functions defined on $S^1\times S^3$, we can absorb the twisting by deforming $S^3$ into three dimensional ellipsoid $S_b^3$, such that we have trivially fibered $\tilde{S}^1$ over $S_b^3$. 
Recall that the $S_b^3$ partition function for three dimensional $\cN=2$ gauge theories computed in \cite{Hama:2011ea} can be readily factorized into two copies of K-theoretic vortex partition functions on $S^1\times_{\rm q} D^2$ \cite{Pasquetti:2011fj, Beem:2012mb} \footnote{One should however note that there are subtleties with the choice of integration contour and charge of the matter fields to ensure factorization. See \cite{Beem:2012mb, Chen:2013pha} for more detailed discussions, also \cite{Fujitsuka:2013fga, Benini:2013yva} for explicit Higgs branch localization computations.}, where the co-dimension two vortex particles now wrap along $S^1$ and correspond to the saddle points in localization calculation. 
This picture helps us to understand choice of superconformal indices. When another $\tilde{S}^1$ is added, we turn the vortex particles into the aforementioned non-perturbative surface defects wrapping on two torus, which should again dominate the 4d superconformal indices. 
{This leads us to relate the residues in superconformal indices and the insertion of surface defects, as precisely done in \cite{Gaiotto:2012xa} for $\cN=2$ superconformal field theories, and further studied in \cite{Alday:2013kda, Bullimore:2014nla}. }
However, it is clear from the prior discussion that we can extend this prescription to $\cN=1$ supersymmetric field theories, as observed recently in demonstrating their factorizability of $\cN=1$ superconformal indices \cite{ Peelaers:2014ima,Yoshida:2014qwa}.
Moreover, thank to the recent progress on computing the elliptic genera or 2d indices for 2d $\cN=(2,2)$ and $\cN=(0,2)$ gauge theories \cite{Gadde:2013dda, Benini:2013nda, Benini:2013xpa, Gadde:2013wq}, which, given the explicit world sheet theories for the surface defects, {enable} us to confirm the expectation by explicitly computing their elliptic genera and matching with the building blocks of the bulk 4d indices. 
The systematic study of heterotic vortices in $\cN=1$ supersymmetric gauge theories mentioned earlier provides us an excellent venue to explore such an idea.
This direction is especially interesting given the wealth of 4d $\cN=1$ IR dualities which have been non-trivially verified through superconformal indices \cite{ Dolan:2008qi, Spiridonov:2009za}. Supposing these 4d dual theories also admit surface defects preserving $\cN=(0,2)$ supersymmetry in their world sheets, we can investigate if there can also be new $\cN=(0,2)$ dualities (or even ``triality'', see \cite{ Gadde:2013lxa}) relating them. 
\paragraph{}
We will continue our introduction in this section, following \cite{Hanany:2004ea, Shifman:2004dr, Edalati:2007vk,Tong:2007qj,Shifman:2008wv, Koroteev:2010ct, Tong:2008qd} to review vortices in 4d $\cN=2$ and $\cN=1$ supersymmetric gauge theories and their world sheet descriptions.
In section \ref{4d-Indices}, we will explicitly compute the superconformal indices for four dimensional $\cN=2$ and $\cN=1$ supersymmetric gauge theories 
related by superpotential deformation.
In particular, we extract the possible contributions from the vortices/surface defects corresponding to the residues in the contour integration. 
In section \ref{2d-Indices}, we compute the elliptic genera for the two dimensional $\cN=(2,2)$ and $\cN=(0,2)$ vortex world sheet theories given in the previous section.
Finally in section \ref{2d4dMatching}, we identify the 4d and 2d fugacity parameters using superconformal algebras, and match precisely the elliptic genera of the vortex theories with the residues of the 4d superconformal indices for arbitrary ranks of gauge and flavor symmetry groups with appropriate R-charge assignment. 
From this we demonstrate that 4d Seiberg duality descend to 2d vortex theory as the invariance under the ``hopping transformation''. 
Moreover, we also consider other $\cN=(0,2)$ surface defects which are unrelated to dynamical vortices, and compute their elliptic genera. Using these data we generalize the Hori-Tong duality for $\cN=(2,2)$ theories \cite{Hori:2006dk} to $\cN=(0,2)$ setting. Meanwhile, combining with the hopping transformation, we also generalize ``triality'' shown in \cite{Gadde:2013dda}.
{In the main text and appendix \ref{Appendix-Factorization}, we discuss the factorization of the 4d superconformal indices, which also helps illustrating that their saddle points naturally correspond to the surface defects, confirming the earlier matching.} 

\subsection{Vortices in Four Dimensional $\cN=2$ Gauge Theory}\label{N=2Vortex}
\paragraph{} 
We begin by considering four dimensional $\cN=2$ supersymmetric gauge theory with $U(N_c)$ gauge group and $N_f$ fundamental flavors. 
Each $\cN=2$ fundamental hypermultiplet can be decomposed into two copies of $\cN=1$ chiral fields denoted as $Q^i_l$ and $\tilde{Q}^l_i$ which transform respectively as $({\bf N}_c, \overline{\bf N}_f)$ and $( \overline{\bf N}_c, {\bf N}_f)$ under the gauge and flavor symmetry group (\ref{N=2Symmetry}).
To such matter contents, we need to couple them with the $\cN=1$ $U(N_c)$ adjoint chiral multiplet in the $\cN=2$ vector multiplet denoted as $A^l_k$:
\begin{equation}\label{N=2SuperPotential}
\cW_{\cN=2}=\sqrt{2}\sum^{N_F}_{i=1}
\tilde{Q}_i(A-\mu_i)Q^i,
\end{equation}
where we have also associated with each flavor a complex mass parameter $\mu_i$. 
We can also regard the complex mass parameters $\{\mu_i\}$ as the diagonal vevs of the adjoint scalar in the background vector multiplet for the flavor symmetry, such that they satisfied $\sum_{i=1}^{N_f} \mu_i = 0$. 
Finally, $U(N_c)$ gauge group contains a non-trivial Abelian factor which allows us to further introduce FI parameter $v^2$, and we denote the four dimensional complexified gauge coupling as:
\be
\tau=\frac{2\pi i}{e^2} + \frac{\theta}{2\pi}.
\ee 
\paragraph{}
In the absence of complex mass parameters $\{\mu_i\}$,
this theory enjoys the following gauge and global symmetries:
\begin{equation}\label{N=2Symmetry}
U(N_c)\times SU(N_f)\times U(1)_J \times U(1)_r.
\end{equation}
The mass parameters break $SU(N_f)$ global symmetry group down to $U(1)^{N_f-1}$,
while the non-manifesting $SU(2)_R$ R-symmetry is broken down to $U(1)_J$ by the FI parameter $v^2$.
Note also that in full quantum theory, in the massless limit, $U(1)_r$  symmetry 
is generically anomalous and broken down to discrete $Z_{(2N_c-N_f)}$ subgroup for $2N_c > N_f$, but  only preserved for the superconformal $N_f=2N_c$ case. 
\footnote{If the complex masses are turned on, the $U(1)_r$ is explicitly broken down to $Z_2$. }
Only $U(1)_J$ which is the Cartan subgroup of the $SU(2)_R$ symmetry remains the non-anomalous R-symmetry. 
\paragraph{}
We will take $N_f>N_c$. In the absence of complex masses $\{\mu_i\}$, we can have  $2N_c(N_f-N_c)$ dimensional Higgs branch of vacua. 
Turning on complex masses such that $\mu_i \gg \nu^2$ lift most of these branches, however we can still have isolated vacuum given by setting $\tQ^l_i =0$
and 
\begin{eqnarray}
&& a^l_k={\rm diag}(\mu_1, \mu_2,\dots, \mu_{N_c}),\label{RootBH}\\
&&Q^i_l = v \delta_l^i, \quad i=1,\dots, N_c; \quad Q_a^i=0, \quad i=N_c+1,\dots, N_f. \label{Higgsvev}
\end{eqnarray}
Here $a^l_k$ is the complex scalar component in  $A^l_k$,
and there are $\frac{N_f!}{(N_f-N_c)! N_c!}$ such isolated vacua, corresponding to selecting $N_c$ out of $N_f$ mass parameters to satisfy (\ref{RootBH}). They are called ``root of Baryonic Higgs branch'', which represent the intersections between Coulomb and Higgs branches.  
This can be seen from the symmetry breaking pattern in (\ref{RootBH}) and (\ref{Higgsvev}):
\begin{equation}\label{SymBreaking}
U(N_c)\times SU(N_f)\xlongrightarrow{\mu_i} U(1)^{N_c} \times U(1)^{N_f-1} \xlongrightarrow{v^2} S[U(1)^{N_c} \times U(1)^{N_f-N_c}].
\end{equation} 
On the other hand if $\nu^2\gg \mu_i$, the symmetry breaking pattern becomes:
\begin{equation}\label{SymBreaking2}
U(N_c)\times SU(N_f)\xlongrightarrow{\nu^2} S[U(N_c)\times U(N_f-N_c)] \xlongrightarrow{\mu_i} S[U(1)^{N_c}\times U(1)^{N_f-N_c}].
\end{equation} 
The condition (\ref{RootBH}) represents special loci on the Coulomb branch vacua, along which there are additional $N_c$ squarks becoming massless and condense to parameterize the remaining Higgs branch. 
\paragraph{}
The symmetry breaking patterns (\ref{SymBreaking}) and (\ref{SymBreaking2}) tell us that the overall $U(1) \subset U(N_c)$ is broken, 
and Higgs branch can support topological vortices charged under each of $N_c$ different diagonal $U(1)$ subgroups, whose topological number can arise from the winding number of the corresponding $Q^i_l$ with vevs $v\delta_l^i$.
The defining equations for vortices oriented along $x_3$ and $x_4$ directions are given by:
\begin{eqnarray}\label{VortexEqn}
&&\frac{1}{e^2}F_{z\bar{z}} = \sum_{i=1}^{N_f} Q^i Q_i^{\dag} -v^2,\nn \\
&& D_{z} Q_l^i = \frac{1}{2}(D_1 - iD_2) Q_l^i =0, \quad \tilde{Q}^l_i = 0, \quad  z=x_1+i x_2.
\end{eqnarray} 
These equations can be obtained from usual Bogomol'nyi completing the square trick, where the solutions are classified by the winding number  $k= {\rm Tr}\int  \frac{F_{12}}{2\pi} \in {\mathbb Z}^{+}$. 
It is interesting to also consider the solution to (\ref{VortexEqn}) in so-called ``singular gauge'' \cite{Edalati:2007vk}, where $Q^i_l$ do not wind asymptotically but the flux is supported by the singular gauge field profile\footnote{Here $r$ is a real parameter which will be identified with the complex combination of two dimensional FI parameter.}:
\begin{equation}\label{singularA}
(A_z)^l_k = -i \bar{z} a(|z|) \left(\frac{\phi^l\bar{\phi}_l}{r}\right) \longrightarrow (A_\vartheta)^l_k = 2|z|^2 a(|z|) \left(\frac{\phi^l\bar{\phi}_k}{r}\right).
\end{equation}
Here $z=|z| e^{i\vartheta}$,  $a(|z|)$ is a radial profile function with asymptotic behavior $a(|z|) \to \frac{1}{2|z|^2},~ |z| \to 0$ and $a(|z|) \to 0,~|z| \to\infty$, and $\phi^l \in {\mathbb C}^{N_c}$ defines the gauge orientation modes. 
This precisely reproduces the desired singular behavior for the gauge field $A \sim \alpha d\vartheta +\dots $ corresponding to surface operator insertions 
at the origin (see for example \cite{Alday:2009fs} ). 
More precisely, surface operators should be regarded as the infra-red limit of the dynamical vortices. However, the four dimensional superconformal indices to be introduced shortly are insensitive to such distinction while these co-dimension two defects will appear as simple poles in the contour integration and give dominant contributions. 
\paragraph{}
The two dimensional vortex world sheet theory preserves half of the four dimensional $\cN=2$ supersymmetry and was first found in \cite{Hanany:2004ea} (see also \cite{Shifman:2004dr}) using explicit D-brane construction.  
We can equivalently quantize and supersymmetrize the zero modes fluctuations around the classical vortex solutions, namely translational and orientational zero modes, plus their fermionic partners (see \cite{Edalati:2007vk} for more details.).
Either way, for winding number $k$, the resultant theory is a $\cN=(2,2)$ $U(k)$ gauge theory with the vector multiplet $U^\alpha_\beta$, plus matter contents of an adjoint chiral multiplet $Z^\alpha_\beta$,  
$N_c$ fundamental chiral multiplets $\Phi^a_\alpha$ and $N_f-N_c$ anti-fundamental chiral multiplets $\tilde{\Phi}^\alpha_j$. 
Finally, we can also combine the two dimensional FI parameter $r > 0$ and theta angle $\vartheta$ into a complex combination:
\begin{equation}
{\bf t} = i r +\frac{\vartheta}{2\pi},
\end{equation}
which can be identified with 4d complex coupling from explicit D-brane construction \cite{Hanany:2004ea}.
We will summarize their transformation properties under the gauge and global symmetry group:
\begin{equation}\label{N=22Symmetry}
U(k)\times SU(N_c)\times SU(N_f-N_c)\times U(1)_c\times U(1)_r\times U(1)_z
\end{equation}
in Table \ref{table:N=(2,2)}. Here $U(1)_r$ and $U(1)_c$ simply descend from the four dimensional abelian global symmetries  (\ref{N=2Symmetry}), 
while $U(1)_z$ arises from the rotational symmetry of vortices in $x^1-x^2$ plane. 
Furthermore, the complex mass parameters $\{\mu_i\}$ in 4d superpotential (\ref{N=2SuperPotential}) become the twisted masses for 
the $\cN=(2,2)$ chiral multiplets, $\Phi^a_\alpha$ and $\tilde{\Phi}_j^\alpha$, and break the flavor symmetry group down to its maximal torus. 
The discrete vacua of the vortex world volume theory is given by
\begin{eqnarray}
&&\sigma^{\alpha}_{\beta}={\rm diag}(\mu_1, \mu_2, \dots, \mu_k),\quad \tilde{\Phi}_j^\alpha = 0, \quad j=N_c+1, \dots N_f \label{N=(2,2)Vac1} \\
&& \Phi^l_\alpha = \sqrt{r} \delta^l_{\alpha}, \quad l =1,\dots, k, \quad \Phi^l_\alpha = 0, \quad l=k, k+1,\dots, N_c.\label{N=(2,2)Vac2}
\end{eqnarray}  
Here $\sigma^\alpha_\beta$ is the scalar component in the $\cN=(2,2)$ $U(k)$ vector multiplet $U^\alpha_\beta$, 
and there are $\frac{N_c!}{(N_c-k)! k!}$ such discrete vacua.
For completeness, we will later compute the elliptic genus of the $\cN=(2,2)$ vortex world volume theory following \cite{Gadde:2013dda}, and we will demonstrate they reproduce identical expression from the residues for the simple poles coming from the hypermultiplets.

\subsection{Vortices in Four Dimensional $\cN=1$ Gauge Theory}\label{N=1Vortex}
\paragraph{}
We can further break four dimensional $\cN=2$ supersymmetry down to $\cN=1$.
One common way is to add a deformation superpotential for the adjoint chiral field $A$ to (\ref{N=2SuperPotential}) to obtain:
\begin{equation}\label{N=1SuperPotential1}
\cW_{\cN=1} =  \sqrt{2}\sum^{N_F}_{i=1}
\tilde{Q}_i(A-\mu_i)Q^i + {\cW}(A) 
\end{equation}
where ${\cW}(A)$ for the time being is an arbitrary holomorphic function for the adjoint scalar in vector multiplet $A$.
Adding ${\cW}(A)$, the Higgs branch solutions get modified into \cite{Edalati:2007vk}:
\begin{eqnarray}
&& 
Q_l^i =  s_i \delta_l^i, \quad \tilde{Q}_i^l = \tilde{s}^i \delta_i^l,\quad \sum_{i=1}^{N_f} |s_i|^2 -|\tilde{s}_i|^2 = v^2,\quad 
i=1, \dots, N_f, \label{N=1eqn1}
\\
&&a_k^l ={\rm diag} (\mu_1,\mu_2,\dots, \mu_{N_c} ),
 \quad \sum_{i=1}^{N_f} s_i \tilde{s}^i  = \frac{\partial {\cW}(x= \mu_l )}{\partial  x }, \quad l, k=1,\dots N_c. \label{N=1eqn2}
\end{eqnarray}  
Furthermore, in order to have topologically stable vortex solutions, it is crucial to impose $\tilde{s}_i = 0$ which restricts the complex mass parameter $\mu_l$ for given $l=1, 2,\dots, N_c$ to coincide with one of the critical points of $\cW(x)$, i. e. 
\begin{equation}\label{N=1condition}
\frac{\partial {\cW}(x = \mu_l)}{\partial x} = 0.
\end{equation} 
This implies that to have non-trivial vortices charged under all $N_c$ $U(1)$ subgroups of $U(N_c)$, 
we need to have 
\begin{equation} \label{4dWmin}
\frac{\partial {\cW}(x)}{\partial x} = \lambda_{N_c} \prod_{l=1}^{N_c}(x-\mu_l)
\end{equation}
which ensures (\ref{N=1eqn2}) is satisfied. 
We can also consider $\hat{\cW}(x)$ of lower degrees, which will allow vortices to be charged only under some of $N_c$ $U(1)$ subgroups.
{We can consider a particular interesting deformation, given by \cite{Argyres:1996eh}} :
\begin{equation}\label{defW2}
{\cW}(A)=\frac{\mathbf{\mu}}{2} {\rm Tr}(A^2),
\end{equation}
and let us also turn off $\{\mu_i\}$ for the time being.
If $\mu >>\Lambda_{\cN=2}$, where $\Lambda_{\cN=2}$ is the $\cN=2$ strong coupling scale, we can integrate out $\Phi$ to obtain an effective potential. 
More precisely, the one loop running of the non-Abelian $SU(N_c) \subset U(N_c)$ gives $\Lambda_{\cN=1}^{3N_c-N_f} = \mu^{N_c}\Lambda^{2N_c-N_f}_{\cN=2}$. 
Therefore, when we take $\mu \to \infty$ and $\Lambda_{\cN=2} \to 0$ while keeping fixed $\Lambda_{\cN=1}$, the physics between $\Lambda_{\cN=1}$  and  $\mu$ can be described by $\cN=1$ SQCD.  
Below $\Lambda_{\cN=1}$, the deformed theory becomes strongly coupled and flows to an interacting infra-red fixed point exhibiting Seiberg duality \cite{Seiberg:1994pq}.
Furthermore, one can restore the complex mass parameters and shift $A^l_k$ to $A^l_k-\mu_k \delta^l_k$ in (\ref{defW2}) to ensure the existence of vortex permitting vacua.
\footnote{The quantum RG flow from $\cN=2$ to $\cN=1$ theory and the moduli space have also been analyzed in details in \cite{Bolognesi:2008sw}.}.  
We will return to this deformation later when we consider the duality between the vortex world volume theories.
\paragraph{}
Moreover, we should note that the $\cN=1$ superpotential explicitly breaks the $U(1)_r\times U(1)_J$ global symmetry inherited from the $\cN=2$ theory.
However, for a given $\hat{\cW}(A)$, it can still preserve a $U(1)$ symmetry from the linear combination of $U(1)_r$ and $U(1)_J$. 
Although, in full quantum theory, such a $U(1)$ symmetry can still suffer chiral anomaly,  if the deformed theory flow to an interacting infra-red fixed point, such as $\cN=1$ SQCD in conformal window discussed earlier, 
$\cN=1$ superconformal symmetry still demands an $U(1)_R$ symmetry to exist, which will arise from the linear combination of global abelian symmetries. 
\paragraph{}
Turning to the vortex world volume theory under such a deformation, from the variations under $\cN=1$ supercharges we can deduce that 
provided the condition (\ref{N=1condition}) is satisfied, the BPS vortex solutions can only be compatible with the chiral $\cN=(0,2)$ supersymmetry \cite{Edalati:2007vk}.
We can also see this from the vortex world volume. Let us decompose the field contents of the previous $\cN=(2,2)$ vortex theory into $\cN=(0,2)$ fields,
such that an $\cN=(2,2)$ vector multiplet is decomposed into an $\cN=(0,2)$ vector multiplet plus an $\cN=(0,2)$ chiral multiplet, and an $\cN=(2,2)$ chiral multiplet is decomposed into an $\cN=(0,2)$ chiral multiplet plus an $\cN=(0,2)$ Fermi multiplet.
Now if we denote the $\cN=(0,2)$ adjoint chiral multiplet which makes up the $\cN=(2,2)$ $U(k)$ vector multiplet as $\Sigma^\alpha_\beta$, it contains $\sigma^\a_\b$ as its scalar component.
The four dimensional superpotential in (\ref{N=1SuperPotential1}) now descends to superpotential $\hat{\cW}(\Sigma)$ for $\Sigma_\beta^\alpha$, and explicitly breaks the world sheet supersymmetry from $\cN=(2,2)$ to $\cN=(0,2)$.
As in the bulk 4d theory, the supersymmetric vacuum of the deformed vortex theory can only exist if we supplement the conditions in (\ref{N=(2,2)Vac1}) and (\ref{N=(2,2)Vac2}) with the following condition:
\begin{equation} \label{2dWmin}
\frac{\partial \hat{\cW}(x = \mu_\alpha)}{\partial x} = 0, 
\end{equation}
for $\alpha$ equals to one of $1, 2, \dots, k$. Clearly the twisted mass parameters $\{\mu_\alpha\}$ are subset of $\{\mu_l\}$ entering \eqref{4dWmin}, which implies that the 4d condition (\ref{4dWmin}) is sufficient to ensure (\ref{2dWmin}) satisfied. 
In Table \ref{table:N=(0,2)}, we summarize the symmetry group and field contents for the resultant $\cN=(0,2)$ vortex world volume theory,
and we will also compute its elliptic genus to demonstrate that we can indeed reproduce it from the $\cN=1$ superconformal index.

\section{4d $\cN=2$ and $\cN=1$ Superconformal Indices}\label{4d-Indices}
\paragraph{}
In this section, we explicitly compute the four dimensional superconformal indices for the $\cN=2$ and $\cN=1$ supersymmetric gauge theories discussed in the previous section, they will be used later to match with the elliptic genera of the vortex world volume theories.
\subsection{4d $\cN=2$ Superconformal Index}
\paragraph{}
The four-dimensional $\cN=2$ superconformal index was introduced in \cite{Kinney:2005ej, Romelsberger:2005eg}, and it is defined to be a twisted partition function on $S^3\times S^1$. 
Such a index counts the states that are annihilated by a chosen supercharge $\cQ=\tilde{\cQ}_{1\dotdiv}$. Following \cite{Gadde:2011ik}, it is defined by:
\begin{equation}\label{DefN=2Index}
\cI^{\cN=2}({\fa}_i; \fp, \fq, \ft)={\rm Tr} (-1)^F \fp^{h_{34}-\fr}\fq^{h_{12}-\fr}\ft^{\fR+\fr} e^{i\beta \delta_{\cN=2}}\prod_i {\fa}_i^{f_i}.
\end{equation}
Here $j_{1,2}$ label the Cartan charges of the isometry group $SU(2)_1\times SU(2)_2$ of $S^3 \subset S^1\times S^3$, such that $h_{12}\equiv j_2+j_1$ and $h_{34}\equiv j_2-j_1$ define respectively the rotational generators on $12$ and $34$ planes, while $(\fR, \fr)$ are the Cartan charges of $SU(2)_R\times U(1)_r$ R-symmetry which are part of $SU(2,2|2)$ superconformal algebra. 
The trace in (\ref{DefN=2Index}) is taken over states of the radially quantized conformal theory on $S^3$,  and only the states satisfying the projection condition:
\begin{equation}\label{defdeltaN=2}
\delta_{\cN=2}\equiv 2\{\mathcal{Q}, \mathcal{Q}^\dagger\}=E-2j_2-2\fR+\fr=0
\end{equation}
can contribute to (\ref{DefN=2Index}), where $E$ denotes the conformal dimension.
In this paper we will also take the fugacity parameters to satisfy 
\begin{equation}\label{FugacityRange}
|\fp|, |\fq|, |\ft|, |\fp\fq/\ft| < 1, \quad |\fa_i| =1.
\end{equation}
As a result, the $\cN=2$ superconformal index is independent of the radius of $S^1$ denoted by $\beta$. 
The complex numbers $\fp$, $\fq$ and $\ft$ are the fugacity parameters for these symmetry generators commuting with $\cQ$ and one another, as well as $\{{\fa}_i\}$, the fugacity parameters for the additional flavor symmetries with the generators $\{f_i\}$.
These fugacity parameters also serve as the regulators truncating the otherwise infinite number of states satisfying the condition (\ref{defdeltaN=2}).
As mentioned earlier, in non-superconformal theory $U(1)_r$  is usually broken to discrete symmetry group, however if the theory flows to infra-red fixed point, 
the anomalous UV $U(1)_r$ needs to mix with the other global abelian symmetries to give rise to the non-anomalous combination at infra-red, as demanded by $\cN=2$ superconfromal algebra. In the actual index computation therefore, $U(1)_r$ charge assignments are treated as free parameters, subsequently fixed to be the non-anomalous values by the anomaly cancelation condition.
\paragraph{}
For an $\cN=2$ theory which admits weakly coupled Lagrangian description represented at UV fixed point that can be driven to an interacting IR fixed point by relevant perturbation along the RG flow, its $\cN=2$ superconformal index can be computed from certain matrix integral whose integrand is determined by the plethystic exponential of so-called ``single particle indices''.  Explicit computations have been done in e.g. \cite{Gadde:2011ik, Gadde:2011uv, Gadde:2013dda}.
For example, for a free hypermultiplet which transform as representation $\Delta$ labeled by a set of fugacity parameters $\{z_{R_j}\}$ under global symmetry group $G$, the index is  
\begin{equation}\label{N=2hyperIndex} 
\cI^{\cN=2}_{H}(z;\fp,\fq,\ft)=
\pl_{R_j\in \Delta}
\pl^\infty_{r,s=0}
\frac{1-z^{\mp}_{R_j}\fp^{r+1}\fq^{s+1}\ft^{-1/2}}{1-z^{\pm}_{R_j} \fp^r \fq^s \ft^{1/2}}=:
\pl_{R_j\in \Delta}
\Gamma\(z^{\pm}_{R_j}\ft^{1/2};\fp,\fq\).
\end{equation}
Notice that in writing down (\ref{N=2hyperIndex}), we have also set the R-charges to be $(\fR, \fr) = (1/2, 0)$ as required by the scalar field in a free theory which has scaling dimension $E=1$. 
Here we have also introduced the notation $\Gamma(x^{\pm}; \fp, \fq) = \Gamma(x; \fp, \fq)\Gamma(x^{-1}; \fp, \fq)$, 
while $\Gamma(x; \fp, \fq)$ is elliptic gamma function defined in appendix \ref{Appendix-Id}, where we also collect most of the definitions and identities for elliptic gamma and theta functions used in the main text.
\paragraph{}
We can also consider the $\cN=2$ vector multiplet contribution to the superconformal index, in particular for the gauge group $U(N_c)$:
\begin{equation}\label{N=2vectorIndex}
\cI_{V}^{\cN=2} = \frac{\ka_4^{N_c}}{N_c!} \oint_{{\mathbb T}^{N_c}} \prod_{l=1}^{N_c}\frac{dz_l}{2\pi i z_l} \frac{\prod_{l,k=1}^{N_c}\Gamma\left(\frac{z_l}{z_k}\frac{\fp\fq}{\ft}; \fp, \fq\right)}{\prod_{l\neq k}^{N_c} \Gamma(\frac{z_l}{z_k}; \fp, \fq)},
\quad
 \kappa_4 = (\fp; \fp)_{\infty} (\fq; \fq)_{\infty} \Gamma({\fp\fq}/{\ft}; \fp,\fq)
\end{equation}
where $\{z_l\}$ are the fugacity parameters for $U(N_c)$ gauge group, 
and the numerator comes from the adjoint chiral multiplet with $(\fR, \fr) = (0, -1)$ such that its scaling dimension again takes the free field value $1$.
Furthermore, $\bbT^{N_c}$ denotes the $N_c$ dimensional unit torus.
To introduce matter contents into pure Yang-Mills theory, we can gauge the flavor and other global symmetries of a free hypermultiplet, and this amounts to insert (\ref{N=2hyperIndex}) into (\ref{N=2vectorIndex}) and perform the integration over the corresponding fugacity parameters.
Combining various contributions,  we can now write down (\ref{DefN=2Index}) for the four dimensional $\cN=2$ supersymmetric gauge theory with $U(N_c)$ gauge theory with $N_f$ fundamental hypermultiplets:
\begin{equation} \label{4DN=2Index}
\cI_{\rm 4D}^{\cN=2} = \frac{\ka_4^{N_c}}{N_c!} \oint_{\bbT^{N_c}} \prod_{l=1}^{N_c}\frac{dz_l}{2\pi i z_l} \frac{\prod_{l,k=1}^{N_c}\Gamma\left(\frac{z_l}{z_k}\frac{\fp\fq}{\ft}; \fp, \fq\right)}{\prod_{l\neq k}^{N_c} \Gamma(\frac{z_l}{z_k}; \fp, \fq)} \prod_{i=1}^{N_f} \prod_{l=1}^{N_c}
\Gamma\left(\(\frac{z_l }{\fa_i}\)^{\pm}\sqrt{\ft}; \fp, \fq\right) 
\end{equation} 
Here $\{z_l\}$ and $\{\fa_i\}$ are fugacities for $U(N_c)$ gauge and $SU(N_f)$ flavor global symmetry, while the $SU(N_f)$ fugacities satisfy $\prod_{i=1}^{N_f} \fa_i =1$, and we can obtain $U(N_c)$ gauge group from $SU(N_c)$ by further gauging $U(1)_B$ global symmetry.
\paragraph{}
Now let us consider the simple poles from the hypermultiplets in (\ref{4DN=2Index}) located at:
\begin{equation}\label{N=2PoleCond}
{z_{l} }  = \fa_{i_l}\ft^{1/2}\fp^{n_{i_l}}\fq^{m_{i_l}} , \quad  l=1, 2, \dots, N_c, \quad n_{i_l}, m_{i_l}\ge 0
\end{equation}
where $1\le i_l \le N_f$ corresponding to selecting $N_c$ out of $N_f$ possible flavor fugacity parameters, and there are $\frac{N_f !}{(N_f-N_c)! N_c!}$ such possible choices. 
This choice of pole condition \eqref{N=2PoleCond} corresponds to picking up the simple poles from the anti-fundamental chirals in the $\cN=2$ hypermultiplet, and they are located inside the unit circle for the range of our fugacity parameters \eqref{FugacityRange}. As we will see later  the residues of these simple poles correspond to the elliptic genus of the dynamical vortex/surface operators on the Higgs branch, for which we expect the results should be symmetrical under exchanging fundamental and anti-fundamental accompanying with changing the sign of FI-parameter. We can implement this by deforming the contour to include the circle at infinity to enclose the simple poles outside the unit circle at ${z_{l} }  = \fa_{i_l}\ft^{-1/2}\fp^{-n_{i_l}}\fq^{-m_{i_l}} $ instead, which correspond to fundamental chirals. We expect the insertion of appropriate FI term contribution into the superconformal index cure the asymptotic behavior at infinity, see \cite{Peelaers:2014ima} for recent discussion on this issue. Indeed, assuming the contributions from the infinity vanishes,  explicitly calculation shows that exchanging the simple poles from fundamental and anti-fundamental amounts to $\fa_l \leftrightarrow 1/\fa_l$ in the final expression to be presented next.
\paragraph{}
When the complex mass parameters $\{\mu_i\}$ vanish, we can regard this condition (\ref{N=2PoleCond}) as generalization of the quantization condition considered in \cite{Dorey:2011pa, Chen:2011sj} for the instanton partition functions in $\Omega$ background.
As discussed in \cite{Gaiotto:2012xa}, the physical interpretation for the simple poles in $\cN=2$ superconformal index indicates certain new flat directions open up, as parameterized by the vevs of tower of new light spatial-dependent baryonic operators and their holomorphic derivatives. This new tower precisely corresponds to the insertion of two distinct sets of surface operators, and the holomorphic derivates with respect to $z=x_1+i x_2$ and $w=x_3+i x_4$ yield the quantized angular momenta carried by them, as indicated by the integer power of fugacity parameters $(\fp, \fq)$ in (\ref{N=2PoleCond}). Furthermore, as both gauge and global symmetries are now partially broken, the residues evaluated at these simple poles compute the index based on the residual superconformal symmetries preserved by these surface defects.  
\paragraph{}
Picking the contour enclosing all the non-negative integral value of $\{n_{i_l}, m_{i_l}\}$ in the contour integration of (\ref{4DN=2Index}), the result is: 
\begin{eqnarray}\label{4DN=2Index-Int}
\cI_{\rm 4D}^{\cN=2} 
&=& 
\frac{\(\Gamma(\frac{\fp\fq}{\ft}; \fp,\fq)\)^{N_c}}{N_c!}
\sum_{\{i_l\}, \{i_k\} \subset \{ i\}} \prod_{\alpha\in\{i_l\}} \prod_{j\in\overline{\{i_{l}\}}}
\frac{\Gamma\(\frac{\fa_{j}} {\fa_\a}; \fp, \fq \)}
{\Gamma\(\frac{\fa_{j}} {\fa_\a}\frac{\fp\fq}{\ft}; \fp, \fq \)}
\sum_{\{n_{i_l}, m_{i_l}\ge 0\}} 
\(\frac{\fp\fq}{\ft}\)^{(N_f-2N_c)\sum_{l=1}^{N_c}n_{i_l}m_{i_l}+2\sum_{l=1}^{N_c}n_{i_l}\sum_{k=1}^{N_c}m_{i_k}}
\nn \\
&\times&
\prod_{\a\in \{i_l\},\b\in\{i_k\}} 
\prod_{r=0}^{n_{\b}-1}
\Delta\(\frac{\fa_{\b}}{\fa_{\a}} \fp^{r-n_{\a}}; \fq, \frac{\fp\fq}{\ft}\) 
\prod_{j\in \overline{\{i_l\}}} 
\prod_{\a \in \{i_l\}} 
\prod_{r=1}^{n_{\a}}
\Delta\(\frac{\fa_j}{\fa_\a}\fp^{-r}; \fq, \frac{\fp\fq}{\ft} \)
\nn\\
&\times&
\prod_{\a\in \{i_l\},\b\in\{i_k\}} 
\prod_{s=0}^{m_{\b}-1}
\Delta\(\frac{\fa_{\b}}{\fa_{\a}} \fq^{s-m_{\a}}; \fp, \frac{\fp\fq}{\ft}\) 
\prod_{j\in \overline{\{i_l\}}} 
\prod_{\a \in \{i_l\}} 
\prod_{s=1}^{m_{\a}} 
\Delta\(\frac{\fa_j}{\fa_a}\fq^{-s}; \fp, \frac{\fp\fq}{\ft} \).
\end{eqnarray} 
Here $\{i_l\}\equiv \{i_l: l=1, 2, \dots N_c\}$, $\{i : i=1, 2,\dots, N_f\}$ and $\overline{\{i_l\}}$ is the complement of subset $\{i_l\}$, and the function $\Delta(x; q, t)$ is the ratio of $\theta(x; q)$ defined in (\ref{def-Delta}), which was first introduced in \cite{Gadde:2013dda}.
We see that the expression above is in almost factorizable form into exclusively $n_{i_l}$ and $m_{i_l}$ dependence, as given by the last two lines above.{ 
In fact, we will explicitly demonstrate later that they correspond to the elliptic genera for two distinct sets of surface defects inserted in orthogonal $x^{1,2}$ and $x^{3,4}$ planes, }
and they interact through the following non-factorizable term in the summation:
\begin{equation}\label{nonfactor1}
\(\frac{\fp\fq}{\ft}\)^{(N_f-2N_c)\sum_{l=1}^{N_c}n_{i_l}m_{i_l}+2\sum_{l=1}^{N_c}n_{i_l}\sum_{k=1}^{N_c}m_{i_k}}.
\end{equation} 
Clearly if either $n_{i_l}=0$ or $m_{i_l}=0$ for all $l=1, 2,\dots, N_c$, the non-factorizable factor above vanishes identically, 
while for generic non-vanishing $\{n_{i_l}\}$ and $\{m_{i_l}\}$ and superconformal limit $N_f=2N_c$, clearly we see that we can have factorizable form if we impose:
\begin{equation}\label{FacCond1}
\sum_{l=1}^{N_c} n_{i_l} = \sum_{l=1}^{N_c} m_{i_l}.
\end{equation}
For $N_f < 2N_c$ we need to in addition impose 
\be\label{FacCond2}
\sum_{l=1}^{N_c} n_{i_l} m_{i_l}=0, 
\ee
i. e. $\{n_{i_l}\}$ and $\{m_{i_l}\}$ are two orthogonal $N_c$ dimensional vectors with non-negative entries, for example, we can have $\{n_{i_l}\}= (1, 0, 1, 0, \dots, 0)$ and $\{m_{i_l}\}=(0, 2, 0, 0,\dots, 0)$.
As individual $n_{i_l}$ and $m_{i_l}$ can be interpreted as the topological charges carried by surface operators under $l$-th $U(1)$ Cartan subgroup of $U(N_c)$, 
we can ensure factorization if they are charged under different $U(1)$s and with equal total topological charges.
Let us now consider the combination where $\{m_{i_l}\}$ vanish, and the remaining summation gives: 
\begin{equation}\label{EGN=2}
\cI^{\cN=2}_{\{n_{i_l}\}, 0}=
\sum_{\{n_{i_l}\ge 0\}} 
\prod_{\a\in \{i_l\},\b\in\{i_k\}} 
\prod_{r=0}^{n_{\b}-1}
\Delta\(\frac{\fa_{\b}}{\fa_{\a}} \fp^{r-n_{\a}}; \fq, \frac{\fp\fq}{\ft}\) 
\prod_{j\in \overline{\{i_l\}}} 
\prod_{\a \in \{i_l\}} 
\prod_{r=1}^{n_{\a}} 
\Delta\(\frac{\fa_j}{\fa_\a}\fp^{-r}; \fq, \frac{\fp\fq}{\ft} \).
\end{equation}
The other one can be obtained by exchanging the fugacity parameters $\fp$ and $\fq$, and $\{n_{i_l}\}$ and $\{m_{i_l}\}$.
We will later demonstrate that (\ref{EGN=2}) precisely coincides with the elliptic genus of the 2d $\cN=(2,2)$ vortex world sheet theory discussed earlier, after the identification of 2d and 4d parameters.   
\paragraph{}
Here we also consider $\cN=2$ quiver gauge theory with gauge group $SU(N_c)\times U(N_c)$ with fugacity parameters $\{z_l\}$ and $\{\tz_l\}$ such that $\prod_{l=1}^{N_c}z_l=1$
and bi-fundamental hypermultiplet joining them, for which the corresponding $\cN=2$ superconformal index is given by:
\begin{eqnarray}\label{N=2Index-Quiver}
\cI^{\cN=2}_{\rm Quiver} &=& \frac{\kappa_4^{N_c-1}}{N_c !}\frac{\kappa_4^{N_c}}{N_c !} \oint _{{\mathbb T}^{N_c-1}} \prod_{l=1}^{N_c-1} \frac{dz_l}{2\pi i z_l}
\frac{\prod_{l, k=1}^{N_c}\Gamma\(\frac{z_l}{z_k}\(\frac{\fp\fq}{\ft}\); \fp, \fq\)}{\prod_{l\neq k}^{N_c}\Gamma\(\frac{z_l}{z_k}; \fp, \fq\)}
\oint _{{\mathbb T}^{N_c}} \prod_{l=1}^{N_c} \frac{d\tz_l}{2\pi i \tz_l}
\frac{\prod_{l, k=1}^{N_c}\Gamma\(\frac{\tz_l}{\tz_k}\(\frac{\fp\fq}{\ft}\); \fp, \fq\)}{\prod_{l\neq k}^{N_c}\Gamma\(\frac{\tz_l}{\tz_k}; \fp, \fq\)}\nn\\
&\times& \prod_{k,l=1}^{N_c}\Gamma\(\(\frac{\tz_l}{z_k}h\)^{\pm}\sqrt{t};\fp,\fq\)\prod_{i=1}^{N_f}\prod_{l=1}^{N_c} \Gamma\(\(\frac{\tfa_j}{\tz_l}\tilde{y}\)^{\pm}\sqrt{t};\fp,\fq\) \Gamma\(\(\frac{z_l}{\fa_j}y\)^{\pm}\sqrt{\ft}; \fp,\fq \).
\end{eqnarray}
Here the first line contains the $\cN=2$ vector multiplet contributions from the $SU(N_c)\times U(N_c)$ quiver gauge group; 
{the second line contains the bi-fundamental hypermultiplets transforming as $({\bf N_c}, \bar{\bf N}_c)$ under $U(N_c)\times SU(N_c)$, }
and additional $N_f$ fundamental and anti-fundamental hypermultiplets charged respectively under $U(N_c)$ and $SU(N_c)$ gauge groups.
Finally the parameters $\{\fa_i\}$ and $\{\tfa_i\}$ correspond to the $SU(N_f)_A\times SU(N_f)_B$ flavor fugacities satisfying $\prod_{i=1}^{N_f}\fa_i = \prod_{i=1}^{N_f} \tfa_i =1$, 
and $h$, $y$ and $\tilde{y}$ are fugacity parameters for $U(1)$ global symmetry rotating individual hypermultiplet.
In this case, we would like to consider the topological vortices arising from Higgsing the $U(N_c)$ gauge group, and this corresponds to considering the simple poles inside the unit circle:
\begin{equation}
\tz_l = z_l h^{-1} t^{1/2} \fp^{n_l} \fq^{m_l}, \quad l=1, 2, \dots, N_c, \quad n_l, m_l \ge 0,
\end{equation} 
and superconformal index (\ref{N=2Index-Quiver}) reduces to
\begin{eqnarray}\label{N=2Index-Quiver-Int}
\cI^{\cN=2}_{\rm Quiver} &=& \frac{\kappa_4^{N_c-1}}{N_c !} \oint _{{\mathbb T}^{N_c-1}} \prod_{l=1}^{N_c-1} \frac{dz_l}{2\pi i z_l}
\frac{\prod_{l, k=1}^{N_c}\Gamma\(\frac{z_l}{z_k}\(\frac{\fp\fq}{\ft}\); \fp, \fq\)}{\prod_{l\neq k}^{N_c}\Gamma\(\frac{z_l}{z_l}; \fp, \fq\)}
\prod_{l=1}^{N_c} \prod_{i=1}^{N_f} \Gamma\(\(\frac{z_l}{\fa_i}y\)^{\pm}\sqrt{\ft}; \fp,\fq \)
\nn\\
&\times& \sum_{\{n_l\}, \{m_l\}} \(\frac{\fp\fq}{\ft}\)^{-N_c \sum_{l=1}^{N_c} n_l m_l +2\sum_{l=1}^{N_c}m_l\sum_{k=1}^{N_c} n_l} \prod_{i=1}^{N_f}\prod_{l=1}^{N_c} \Gamma\(\(\frac{\ta_i}{z_l} (h t^{-1/2} \tilde{y}) \fp^{-n_l}\fq^{-m_l}\)^{\pm}\sqrt{t}; \fp,\fq\)\nn\\
&\times& \prod_{l,k=1}^{N_c}\left[ \prod_{r=0}^{n_k-1} \Delta\(\frac{z_k}{z_l}\fp^{r-n_l};\fq,\(\frac{\fp\fq}{\ft}\)\) 
\prod_{s=0}^{m_k-1} \Delta\(\frac{z_k}{z_l}\fq^{s-m_l};\fp,\(\frac{\fp\fq}{\ft}\)\)\right].
\end{eqnarray}
If we set all $\{m_l\}$ vanish, $N_f=N_c$ and $x=ht^{-1/2}\tilde{y}$, we recover the equation (6.4) of \cite{Gadde:2013dda} for the insertion of surface operators into $\cN=2$ superconformal field theory, through such systematic higgsing of quiver gauge group.
This is the first evidence that the residues in superconformal index precisely yield the elliptic genus of the corresponding surface defects.

\subsection{4d $\cN=1$ Superconformal Index}
\paragraph{}
We can also analogously introduce four dimensional $\cN=1$ superconformal index from the $SU(2,2|1)$ superconformal algebra, which is given by \cite{Gadde:2010en}:
\begin{equation}\label{defN=1Index}
\cI^{\cN=1}(\fa_i; \fp,\fq)=
{\rm Tr}(-1)^F \fp^{h_{34}+\frac{\tilde{\fr}}{2}}
\fq^{h_{12}+\frac{\tilde{\fr}}{2}} e^{i\beta \delta_{\cN=1}}
\prod_i \fa_i^{f_i},
\end{equation}
such that the index only counts the protected states satisfying the projection condition:
\begin{equation}\label{defdeltaN=1}
\delta_{\cN=1}=\{\mathcal{Q},\mathcal{Q}^\dagger\}=E-2j_2-\frac{3}{2}\tilde{\fr}=0,
\end{equation} 
where $\tilde{\fr}$ denotes the $\cN=1$ R-charge. One should note that while in free UV fixed point we can assign R-charge to free field to be $2/3$ so that it has scaling dimension $E=1$, the $U(1)_R$ charge denoted as $\tilde{\fr}$ entering (\ref{defN=1Index}) should be the non-anomalous one appearing at the infra-red fixed point. The $U(1)_R \subset SU(2,2|1)$ appearing in the IR fixed point arises from mixing with other global symmetries, i. e. 
\begin{equation}
\tilde{\fr} = \tilde{\fr}_0 + \sum_{i} \tilde{s}_i  f_i,
\end{equation} 
where $\tilde{\fr}_0$ is the UV $U(1)_R$ charge and  the real coefficients $\{s_i\}$ are determined through a-maximization principle \cite{Intriligator:2003jj}.
Substituting back into (\ref{defN=1Index}), we obtain  
\begin{equation}\label{defN=1Index2}
\cI^{\cN=1}(\fa_i; \fp,\fq)=
{\rm Tr}(-1)^F \fp^{h_{34}+\frac{\tilde{\fr}_0}{2}}
q^{h_{12}+\frac{\tilde{\fr}_0}{2}} 
\prod_i (\fa_i (\fp\fq)^{\frac{\tilde{s}_i}{2}})^{f_i},
\end{equation}
so the computation for the superconformal index can still be performed with UV R-charge assignments when the definition of flavor fugacity parameters are shifted as ${\fa}_i \to {\fa}_i (\fp\fq)^{s_i/2}$. See \cite{Beem:2012yn} for more detailed discussion on this issue.
\paragraph{}
More explicitly, for an $\cN=1$ chiral multiplet transforming as the representation $\Delta$ labeled by fugacity parameters $\{z_{R_j}\}$ under symmetry group $G$ and carries $U(1)_R$ charge $\tilde{\fr}$, its contribution to (\ref{defN=1Index}) is given by:
\begin{equation}\label{N=1Index-Chiral}
\cI^{\cN=1}_{\chi}(z;\fp,\fq)=
\pl_{R_j\in \Delta} 
\Gamma\((\fp\fq)^{\frac{\tilde{\fr}}{2}}z_{R_j};\fp,\fq\).
\end{equation}
From the inversion identity for elliptic gamma function (\ref{Inversion-Id}),
we can also deduce that if the superpotential contains complex mass term  $\sim \mu_i \tQ_i Q^i $ for a pair of $\cN=1$ fundamental and anti-fundamental chiral multiplets, their contributions precisely cancel each other.
Finally we can also write down the $\cN=1$ vector multiplet contribution to (\ref{defN=1Index}) for $U(N_c)$ gauge group:
\begin{eqnarray}
\cI_{V}^{\cN=1} &=& \frac{(\fp;\fp)^{N_c}_\infty(\fq;\fq)^{N_c}_\infty}{N_c!} \oint_{\bbT^{N_c}} \prod_{l=1}^{N_c}
\frac{dz_l}{2\pi i z_l} 
\pl^{N_c}_{l\neq k} \frac{1}{\Gamma(\frac{z_l}{z_k}; \fp, \fq)}\nn \\
 &=&\frac{(\fp;\fp)^{N_c}_\infty(\fq;\fq)^{N_c}_\infty}{N_c!} \oint_{\bbT^{N_c}} \prod_{l=1}^{N_c}
\frac{dz_l}{2\pi i z_l} 
\pl^{N_c}_{l> k} \theta({z_l}/{z_k}; \fq)\theta({z_k}/{z_l}; \fp),
\end{eqnarray}
where the theta function is defined in (\ref{def-theta})
and we have used the identity:
\begin{equation}
\Gamma(x; \fp, \fq) \Gamma(x^{-1}; \fp, \fq) = \frac{1}{\theta(x; \fp)\theta(x^{-1}; \fq)}.
\end{equation}
Directly combining all contributions, we can evaluate the $\cN=1$ superconformal index for various theories.
\paragraph{}
As we also consider breaking four dimensional $\cN=2$ supersymmetry to $\cN=1$ supersymmetry via superpotential $\cW(A)$, 
we would like to relate the $\cN=1$ and $\cN=2$ superconformal indices (\ref{defN=1Index}) and (\ref{DefN=2Index}) by identifying their parameters. 
First we notice that from the projection conditions (\ref{defdeltaN=2}) and (\ref{defdeltaN=1})
\footnote{As explained in \cite{Gadde:2010en} there are in fact two inequivalent choices for the $\cN=1$ superconformal choices called ``left-handed'' and ``right-handed", corresponding to the different choices of supercharges and counting only anti-chiral and chiral multiplet contribution. Here we selected the right-handed index in (\ref{defN=1Index}) in order to match the choice of supercharge made in (\ref{DefN=2Index}).}, for the two different indices to count the same set of states, we need to identify their R-charges as:
\begin{equation}\label{RchargeMap}
\tilde{\fr}=\frac{2}{3}(2\fR-\fr).
\end{equation}
Next let us consider the superconformal limit $N_f=2N_c$ and $\mu_i=0$ in (\ref{4DN=2Index}), and add the relevant mass term deformation (\ref{defW2}) to such a UV fixed point. This drives the theory to $\cN=1$ SQCD but now with an additional quartic superpotential $\sim \mu^{-1}Q\tilde{Q} Q\tilde{Q}$ after integrating out the adjoint field $A^a_b$. 
As noted in \cite{Gadde:2010en} that if we set $\ft=(\fp\fq)^{1/2}$ in the $\cN=2$ superconformal index, {the adjoint chiral multiplet contribution in the numerator now becomes unity under the inversion identity (\ref{Inversion-Id})}, while the remaining hypermultiplet becomes a pair of $\cN=1$ fundamental and anti-fundamental chiral multiplets with $U(1)_R$-charge $\tilde{\fr}_A=1/2$. 
Notice this reflects the quartic superpotential being marginal, and it is also non-anomalous charge assignment at the IR fixed point which differs from the UV free field R-charge value of $2/3$.
\paragraph{}
{We propose that this reduction should be a special case of the general deformation potential given by $\cW(A) = \frac{\mu_{K+1}}{K+1}{\rm Tr}( A^{K+1})$, for which the non-anomalous $U(1)_R$ charge assignment at IR fixed point can be recovered by setting:}
\begin{equation}\label{BreakingCond}
\ft = (\fp\fq)^{\frac{K}{K+1}}.  
\end{equation}
The numerator in superconformal index (\ref{4DN=2Index}) now reduces instead to that of an $\cN=1$ adjoint chiral multiplet with $U(1)_R$ charge $\tilde{\fr}_{\rm A}=\frac{2}{K+1}$, while the remaining hypermultiplets become pairs of $\cN=1$ fundamental and anti-fundamental chiral multiplet with $\tilde{\fr}_f=\frac{K}{K+1}$. 
For $N_f=2N_c$, these $U(1)_R$ charge assignments are again non-anomalous, and interestingly the resultant expression precisely corresponds to the $\cN=1$ superconformal index for the electric theory of Kutasov-Schwimmer duality with $N_f=2N_c$ \cite{Kutasov:1995np}.
For $N_f < 2N_c$,  the effective theory obtained from integrating out $A^l_k$ only serves the intermediate step in the RG flow, 
and the R-charge assigned to fundamental and anti-fundamental chiral multiplets obtained from imposing (\ref{BreakingCond}) can be regarded as $\tilde{\fr}_0$ which can mix with flavor symmetries, while the R-charge for the integrated out $A^l_k$ is freezes at $\tilde{\fr}_{\rm A}=\frac{2}{K+1}$. 
As the theory flows further to the deep infra-red fixed point, {the fundamental and anti-fundamental chiral pick up the non-anomalous $U(1)_R$ charge $\tilde{\fr}_{\rm f}=1-\frac{2}{K+1}\frac{N_c}{N_f}$, for which we recover the case for mass deformation when $K=1$.}
\paragraph{}
Let us now write down the $\cN=1$ superconformal index for the theory deformed by the superpotential ${\cW}(A)$:
\begin{eqnarray}\label{N=1IndexElectric}
I^{\cN=1}_{\rm 4D}&=&
\frac{(\fp;\fp)^{N_c}_\infty(\fq;\fq)^{N_c}_\infty}{N_c !}
\oint_{\bbT^{N_c}}\pl^{N_c}_{l=1}
\frac{dz_l}{2\pi iz_l}
\frac{\pl^{N_c}_{l, k=1}\Gamma\((\fp\fq)^{\frac{\tfr_{\rm A}}{2}}\frac{z_l}{z_k};\fp,\fq\)}
       {\pl^{N_c}_{l\neq k}\Gamma\(\frac{z_l}{z_k};\fp,\fq\)}
\pl^{N_F}_{i=1}\pl^{N_c}_{l=1}
\Gamma\(\frac{z_l}{\fa_i} (\fp\fq)^{\frac{\tfr_{\rm f}}{2}};\fp,\fq\)
\Gamma\(\frac{ \fa_i}{z_l}(\fp\fq)^{\frac{\tfr_{\rm af}}{2}};\fp,\fq\).\nn\\
\end{eqnarray}
The $U(1)_R$ charge assignment here need to satisfy:
\begin{equation}\label{RChargeCond}
\tfr_{\rm A} + \tfr_{\rm f} +\tfr_{\rm af} = 2.
\end{equation}
We can interpret this condition on $U(1)_R$ charges of different $\cN=1$ chiral fields as being preserved by the explicit UV $\cN=2$ superpotential (\ref{N=2SuperPotential}), while for $N_f=2N_c$ the condition (\ref{RChargeCond}) is also precisely the condition for the $U(1)_R$ charge assignment to be non-anomalous at the IR fixed point.
To perform the contour integration, as in the 4d $\cN=2$ case we consider the simple poles at the position:
\begin{equation}
z_l = \fa_{i_l} (\fp\fq)^{\tfr_{\rm af}/2} \fp^{n_{i_l}} \fq^{m_{i_l}}, \quad n_{i_l}, m_{i_l} \ge 0.
\end{equation}
The final expression is given by 
\begin{eqnarray}\label{4DN=1Index-Int}
\cI_{\rm 4D}^{\cN=1} 
&=& 
\sum_{\{i_l\}, \{i_k\} \subset \{ i\}} \prod_{\alpha\in\{i_l\}} \prod_{j\in\overline{\{i_{l}\}}}
\frac{\Gamma\(\frac{\fa_{j}} {\fa_\a}; \fp, \fq \)}
{\Gamma\(\frac{\fa_{j}} {\fa_\a}(\fp\fq)^{\tfr_{\rm A}/2}; \fp, \fq \)}
\sum_{\{n_{i_l}, m_{i_l}\ge 0\}} 
\left[\({\fp\fq}\)^{\tfr_{\rm A}/2}\right]^{(N_f-2N_c)\sum_{l=1}^{N_c}n_{i_l}m_{i_l}+2\sum_{l=1}^{N_c} n_{i_l}\sum^{N_c}_{k=1}m_{i_l}}
\nn \\
&\times&
\prod_{\a\in \{i_l\},\b\in\{i_k\}} 
\prod_{j\in \overline{\{i_l\}}} 
 \left[
 \prod_{r=0}^{n_\b -1}
\frac{\theta\(\frac{a_{\b}}{a_{\a}}(\fp\fq)^{\tfr_{\rm A}/2}\fp^{r-n_\a};\fq\)}
       { \theta\(\frac{a_{\b}}{a_{\a}}\fp^{r-n_\a};\fq\) }
\prod_{r=1}^{n_\a} 
\frac{\theta\(\frac{a_{j}}{a_{\a}}(\fp\fq)^{\tfr_{\rm A}/2}\fp^{-r};\fq\)}
{ \theta\(\frac{a_{j}}{a_{\a}}\fp^{-r};\fq\) }\right]
\nn\\
&\times&
\prod_{\a\in \{i_l\},\b\in\{i_k\}}
\prod_{j\in \overline{\{i_l\}}} 
 \left[
 \prod_{r=0}^{m_\b -1}
\frac{\theta\(\frac{a_{\b}}{a_{\a}}(\fp\fq)^{\tfr_{\rm A}/2}\fq^{r-m_\a};\fp\)}
       { \theta\(\frac{a_{\b}}{a_{\a}}\fq^{r-m_{\a}};\fp\) }
\prod_{r=1}^{m_\a} 
\frac{\theta\(\frac{a_{j}}{a_{\a}}(\fp\fq)^{\tfr_{\rm A}/2}\fq^{-r};\fp\)}{ \theta\(\frac{a_{j}}{a_{\a}}\fq^{-r};\fp\) }\right]
\end{eqnarray} 
which can also be recovered directly from (\ref{4DN=2Index-Int}) by setting $\ft=(\fp\fq)^{1-\tfr_{A}/2}$, and we notice that the factorization conditions are the same as given in (\ref{FacCond1}) and (\ref{FacCond2}).
In the next section, we will also compute explicitly the elliptic genus for the $\cN=(0,2)$ vortex world sheet theory obtained from the deformation of $\cN=(2,2)$ theory described earlier \cite{Edalati:2007vk, Tong:2007qj}, and demonstrate that it precisely coincides with the second and third lines above after identifications of parameters.
\paragraph{}
Notice that if we relax the R-charge constraint \eqref{RChargeCond} from the $\cN=2$ superpotential, the superconformal index (\ref{N=1IndexElectric}) becomes the one used to check Seiberg \cite{Seiberg:1994pq} or Kutasov-Schwimmer \cite{Kutasov:1995np} dualities, after assigning the anomaly-free R-charges. The integrated expression now becomes:
\begin{eqnarray}\label{4DN=1Index-Int2}
&&{\cI}_{\rm E}^{\cN=1} 
= 
\sum_{\{i_l\}, \{i_k\} \subset \{ i\}} \prod_{\alpha\in\{i_l\}, \beta\in\{i_k\}} \prod_{j\in\overline{\{i_{l}\}}}
\frac{\Gamma\(\frac{\fa_\b}{\fa_\a}(\fp\fq)^{\frac{\tfr_{\rm A}}{2}};\fp,\fq\)}{\Gamma\(\frac{\fa_{\b}} {\fa_\a}(\fp\fq)^{1-\frac{(\tfr_{\rm f}+\tfr_{\rm af})}{2}}; \fp, \fq \)}
\frac{\Gamma\(\frac{\fa_{j}} {\fa_\a}; \fp, \fq \)}
{\Gamma\(\frac{\fa_{j}} {\fa_\a}(\fp\fq)^{1-\frac{(\tfr_{\rm f}+\tfr_{\rm af})}{2}}; \fp, \fq \)}
\nn\\
&&\sum_{\{n_{i_l}, m_{i_l}\ge 0\}} 
\({\fp\fq}\)^{\left[N_f -\frac{N_f(\tfr_{\rm f}+\tfr_{\rm af})}{2} -N_c\tfr_{\rm A} \right]\sum_{l=1}^{N_c}n_{i_l}m_{i_l}+\tfr_{\rm A}\sum_{l=1}^{N_c} n_{i_l}\sum^{N_c}_{k=1}m_{i_l}}
\nn \\
&\times&
\prod_{\a\in \{i_l\},\b\in\{i_k\}} \prod_{j\in \overline{\{i_l\}}} 
\prod_{r=0}^{n_{\beta}-1} \frac{ \theta\(\frac{\fa_{\b}}{\fa_{\a}}(\fp\fq)^{\frac{\tfr_{\rm A}}{2}}\fp^{r-n_{\alpha}};\fq\)
}
{ \theta\(\frac{\fa_{\b}}{\fa_{\a}}\fp^{r-n_{\alpha}};\fq\) }
\prod_{r=1}^{n_{\alpha}} 
\frac{\theta\(\frac{\fa_{j}}{\fa_{\a}}(\fp\fq)^{1-\frac{\tfr_{\rm f}+\tfr_{\rm af}}{2}}\fp^{-r};\fq\)}{ \theta\(\frac{\fa_{j}}{\fa_{\a}}\fp^{-r};\fq\) }
\prod_{r= 1}^{n_{\alpha}}\frac{\theta\(\frac{\fa_{\b}}{\fa_{\a}}(\fp\fq)^{1-\frac{(\tfr_{\rm f}+\tfr_{\rm af})}{2}}\fp^{-r};\fq\)}
{ \theta\(\frac{\fa_{\b}}{\fa_{\a}}(\fp\fq)^{\frac{\tfr_{\rm A}}{2}}\fp^{-r};\fq\)  }
\nn\\
&\times&
\prod_{\a\in \{i_l\},\b\in\{i_k\}} \prod_{j\in \overline{\{ i_l\}}}
\prod_{s=0}^{m_{\beta}-1} \frac{ \theta\(\frac{\fa_{\b}}{\fa_{\a}}(\fp\fq)^{\frac{\tfr_{\rm A}}{2}}\fq^{s-m_{\alpha}};\fp\)
}
{ \theta\(\frac{\fa_{\b}}{\fa_{\a}}\fq^{s-m_{\alpha}};\fp\) }
\prod_{s=1}^{m_{\alpha}} 
\frac{\theta\(\frac{\fa_{j}}{\fa_{\a}}(\fp\fq)^{1-\frac{\tfr_{\rm f}+\tfr_{\rm af}}{2}}\fq^{-s};\fp\)}{ \theta\(\frac{\fa_{j}}{\fa_{\a}}\fq^{-s};\fp\) }
\prod_{s=1}^{m_\a}\frac{ \theta\(\frac{\fa_{\b}}{\fa_{\a}}(\fp\fq)^{1-\frac{(\tfr_{\rm f}+\tfr_{\rm af})}{2}}\fq^{-s};\fp\)}
{ \theta\(\frac{\fa_{\b}}{\fa_{\a}}(\fp\fq)^{\frac{\tfr_{\rm A}}{2}}\fq^{-s};\fp\)
 }
.\nn\\
\end{eqnarray} 
{In this case, we notice that the factorization can be achieved if, in addition to (\ref{FacCond1}), we have:}
\begin{equation}\label{FacCond3}
N_f =\frac{N_f(\tfr_{\rm f}+\tfr_{\rm af})}{2} +N_c\tfr_{\rm A}
\end{equation}
which is precisely the $U(1)_R$ anomaly cancelation condition\footnote{Such a factorization conditions were also noticed in \cite{Yoshida:2014qwa, Peelaers:2014ima}}. 
Here the subscript ``${\rm E}$'' on the index stands for ``Electric Theory'',
such that if we set $\tfr_{\rm A}=\frac{2}{K+1}$ and $\tfr_{\rm f}=\tfr_{\rm af} = 1-\frac{2}{(K+1)}\frac{N_c}{N_f}$, the relevant expression for verifying the duality in \cite{Kutasov:1995np} is recovered,
and  for $K=1$, we recover that for Seiberg duality given in \cite{Seiberg:1994pq}.
For later purpose, we also consider the $\cN=1$ superconformal index for the magnetic dual theory by making the following changes in (\ref{N=1IndexElectric}):
\begin{equation}
N_c \to \tilde{N_c}= KN_f - N_c, \quad \fa_i \to {1}/{\fa_i}, \quad \tfr_{\rm f}, \tfr_{\rm af} = 1-\frac{2}{(K+1)}\frac{N_c}{N_f} \to  1-\frac{2}{(K+1)}\frac{\tilde{N}_c}{N_f},
\end{equation}  
also adding the contributions from additional $K$ gauge-singlet mesonic chiral fields $M_{\iota}, ~{\iota} =1, \dots, K$ of $U(1)_R$ charges $\tfr_{\iota}'=2-\frac{4N_c}{(K+1)N_f}+\frac{2({\iota}-1)}{(K+1)}$. The equalities for the electric and magnetic superconformal indices in integral forms for the non-anomalous R-charge assignments have been proved elegantly in \cite{Dolan:2008qi, Spiridonov:2009za}. Here we write down the integrated expression for the magnetic superconformal index:
\begin{eqnarray}\label{4DN=1Index-Int3}
&&{\cI}_{\rm M}^{\cN=1} 
= 
\sum_{\{i_l\}, \{i_k\} \subset \{ i\}} \prod_{\alpha\in\{i_l\}, \beta\in\{i_k\}} \prod_{j\in\overline{\{i_{l}\}}}
\frac{\Gamma\(\frac{\fa_\a}{\fa_\b}(\fp\fq)^{\frac{\tfr_{\rm A}'}{2}}\)\Gamma\(\frac{\fa_{\a}} {\fa_j} \)}
{\Gamma\(\frac{\fa_{\a}} {\fa_\b}(\fp\fq)^{1-\frac{(\tfr_{\rm f}'+\tfr_{\rm af}')}{2}} \) \Gamma\(\frac{\fa_{\a}} {\fa_j}(\fp\fq)^{1-\frac{(\tfr_{\rm f}'+\tfr_{\rm af}')}{2}} \)  }
\prod_{\iota=1}^K \prod_{i, j=1}^{N_f} \Gamma\(\frac{\fa_i}{\fa_j}(\fp\fq)^{\tfr_{\iota}'}\)
\nn\\
&&\sum_{\{n_{i_l}', m_{i_l}'\ge 0\}} 
\({\fp\fq}\)^{\left[N_f -\frac{N_f(\tfr_{\rm f}'+\tfr_{\rm af}')}{2} -\tN_c\tfr_{\rm A} \right]\sum_{l=1}^{\tN_c}n_{i_l}'m_{i_l}'+\tfr_{\rm A}\sum_{l=1}^{\tN_c} n_{i_l}'\sum^{\tN_c}_{k=1}m_{i_l}'}
\nn \\
&\times&
\prod_{\a\in \{i_l\},\b\in\{i_k\}} \prod_{j\in \overline{\{i_l\}}} 
\prod_{r=0}^{n_{\beta}'-1}\frac{ \theta\(\frac{\fa_{\a}}{\fa_{\b}}(\fp\fq)^{\frac{\tfr_{\rm A}}{2}}\fp^{r-n_{\alpha}'};\fq\)
}
{ \theta\(\frac{\fa_{\a}}{\fa_{\b}}\fp^{r-n_{\alpha}'};\fq\) }
\prod_{r=1}^{n_{\alpha}'} 
\frac{\theta\(\frac{\fa_{\a}}{\fa_{j}}(\fp\fq)^{1-\frac{\tfr_{\rm f}'+\tfr_{\rm af}'}{2}}\fp^{-r};\fq\)}{ \theta\(\frac{\fa_{\a}}{\fa_{j}}\fp^{-r};\fq\) }
\prod_{r=1}^{n_\a'}\frac{\theta\(\frac{\fa_{\a}}{\fa_{\b}}(\fp\fq)^{1-\frac{(\tfr_{\rm f}'+\tfr_{\rm af}')}{2}}\fp^{-r};\fq\)}
{\theta\(\frac{\fa_{\a}}{\fa_{\b}}(\fp\fq)^{\frac{\tfr_{\rm A}}{2}}\fp^{-r};\fq\)}
\nn\\
&\times&
\prod_{\a\in \{i_l\},\b\in\{i_k\}} \prod_{j\in \overline{\{i_l\}}}
\prod_{s=0}^{m_{\beta}'-1}\frac{ \theta\(\frac{\fa_{\a}}{\fa_{\b}}(\fp\fq)^{\frac{\tfr_{\rm A}}{2}}\fq^{s-m_{\alpha}'};\fp\)}
{ \theta\(\frac{\fa_{\a}}{\fa_{\b}}\fq^{s-m_{\alpha}'};\fp\) }
\prod_{s=1}^{m_{\alpha}'} 
\frac{\theta\(\frac{\fa_{\a}}{\fa_{j}}(\fp\fq)^{1-\frac{\tfr_{\rm f}'+\tfr_{\rm af}'}{2}}\fq^{-s};\fp\)}{ \theta\(\frac{\fa_{\a}}{\fa_{j}}\fq^{-s};\fp\) }
\prod_{s=1}^{m_\a'}\frac{\theta\(\frac{\fa_{\a}}{\fa_{\b}}(\fp\fq)^{1-\frac{(\tfr_{\rm f}'+\tfr_{\rm af}')}{2}}\fq^{-s};\fp\)}
{ \theta\(\frac{\fa_{\a}}{\fa_{\b}}(\fp\fq)^{\frac{\tfr_{\rm A}}{2}}\fq^{-s};\fp\) } 
.\nn\\
\end{eqnarray} 
Here $l, k=1, \dots, N_f-N_c$, $n'_{i_l}, m_{i_l}' \ge 0$ and we have used the shorthand notation $\Gamma(x)=\Gamma(x;\fp,\fq)$ in the expression above.
Clearly the residues in the last two lines in (\ref{4DN=1Index-Int2}) and (\ref{4DN=1Index-Int3}) share identical functional expression after exchanging $N_c \leftrightarrow N_f-N_c$ 
,$\{n_\a, m_\a\} \leftrightarrow \{n_\a', m_\a'\}$ and $(\tfr_{\rm f}, \tfr_{\rm af})\leftrightarrow (\tfr_{\rm f}', \tfr_{\rm af}')$. 

\section{2d $\cN=(2,2)$ and $\cN=(0,2)$ Elliptic Genera for Vortex World Sheet Theories}\label{2d-Indices}
\paragraph{}
Let us now introduce the superconformal indices for 2d $\cN=(2,2)$ and $\cN=(0,2)$ gauge theories, which are twisted partition function on $T^2 \equiv S^1\times S^1$ or ``elliptic genus''.
Beginning with $\cN=(2,2)$ depending on the boundary conditions, we can define this quantity in two different ways such that it counts only NSNS or RR sector.
However, if the theory considered flows to an IR superconformal fixed point, we expect the two definitions can be related up to a pre-factor via spectral flow. 
Here we follow \cite{Gadde:2013dda, Gadde:2013wq} to choose right moving supercharge $\cG\equiv \cG_R^{-}$ in the $\cN=(2,2)$ superconformal algebra so that the elliptic genus only counts the NSNS sector:
\begin{equation}\label{defN=(2,2)Index}
\cI^{\cN=(2,2)}({\bf a}_i;q,y)={\rm Tr}(-1)^Fq^{H_{\rm L}}y^{J_{\rm L}}
\pl_i {\bf a}_i^{f_i},
\end{equation}
where the $H_{\rm L}$ and $J_{\rm L}$ denote the left-moving conformal dimension and left-moving $U(1)_R$-symmetry generator, commuting with a chosen supercharge $\mathcal{G}$. 
The parameter $q= e^{2\pi i \hat{\tau}}$ where $\hat{\tau}=\hat{\tau}_1+i\hat{\tau}_2$ is the complex structure modulus of $T^2$, while $y$ is the fugacity parameter for $J_{\rm L}$, and $\{\ba_i\}$ are the fugacity parameters for the remaining flavor symmetries.
The trace is taken over the states satisfying the condition:
\begin{equation}\label{N=(2,2)Cond}
\delta^{\cN=(2,2)}=\{\mathcal{G},\mathcal{G}^\dagger\}=2H_{\rm R}-J_{\rm R}=0,
\end{equation}
where $H_{\rm R}$ and $J_{\rm R}$ are respectively right-moving counterparts of $H_{\rm L}$ and $J_{\rm L}$. 
The choice of $\cG=\cG_R^{-}$ in defining (\ref{defN=(2,2)Index}) will be justified when we consider how $\cN=(2,2)$ and $\cN=(0,2)$ superconformal algebras can be embedding into those of 4d $\cN=2$ and $\cN=1$ respectively.
Notice that at the fixed point, $\cN=(2,2)$ superconformal algebra dictates that there should be two non-anomalous $U(1)$  R-symmetries, and they should again come from the linear combinations of UV R-symmetries and other flavor symmetries.
\paragraph{}
We can analogously define the elliptic genus for the 2d $\cN=(0,2)$ gauge theory, preserving only the the right-moving supersymmetries, which is enhanced to $\cN=(0,2)$ superconformal symmetries when the theory flows to an infra-red fixed point.
The main distinction now is that the R-symmetry is only present in the right-moving sector, so the definition for superconformal index becomes
\begin{equation}\label{DefN=(0,2)Index}
\cI^{\cN=(0,2)}({\bf a}_i;q)
={\rm Tr}(-1)^F q^{H_L}\pl_i{\bf a}_i^{f_i},
\end{equation}
where the $H_L$ is still the left-moving conformal dimension which commutes with the supercharge, and the index only counts the states satisfying:
\begin{equation}\label{N=(0,2)Cond}
\delta^{\cN=(0,2)}=\{\mathcal{G},\mathcal{G}^\dagger\}=2H_R-J_R=0. 
\end{equation}
{In next section, we will embed 2d $\cN=(2,2)$ or $\cN=(0,2)$ superconformal algebra respectively into 4d $\cN=2$ or $\cN=1$ superconformal algebras,
which allows us to precisely identify the fugacity parameters for the 2d and 4d theories and demonstrate the residue of 4d superconformal indices indeed reduce to the elliptic genera of 2d vortex theories.}

\subsection{$\cN=(2,2)$ and $\cN=(0,2)$ Vortex Partition Functions} 
\paragraph{}
Let us now compute the elliptic genus for the $\cN=(2,2)$ k-vortex world sheet theory considered in section \ref{N=2Vortex}. We first summarize below the transformation properties of the matter contents under the  gauge and global symmetry group in Table \ref{table:N=(2,2)}.
\begin{table}[ht]
\centering
\begin{tabular}{|l|c|c|c|}
\hline  
& $\Phi$ & $\tilde{\Phi}$ &$Z$\\\hline  
$U(k)_G$ &$\rm\bf k$&$\bar{\rm\bf k}$&$\rm\bf adj.$\\\hline
$SU(N_c)_A$  & ${{\rm \overline{\bf {N}}_c}}$& {\bf 1}& {\bf 1} \\\hline
$SU(N_F-N_c)_B $ & {\bf 1} & ${\rm\bf  {N_F-N_c }}$ & {\bf 1}\\\hline
$U(1)_c$&1&1&0\\\hline
$U(1)_r$&$R_{\rm f}$&$R_{\rm af}$&$-1$\\\hline
$U(1)_z$&$0$&$0$&$-1$\\\hline
\end{tabular}
\caption{Field contents of $\cN=(2,2)$ vortex world volume theory}
\label{table:N=(2,2)}
\end{table}
The elliptic genus for $\cN=(2,2)$ $U(k)$ gauge theory with these matter contents is given by
\footnote{The special case of $N_f=2N_c$ has been computed in \cite{Gadde:2013dda}}: 
\begin{eqnarray}\label{N=(2,2)Index}
\cI^{\cN=(2,2)}&=&
\frac{\kappa_2^k}{k!}
\oint\pl^k_{\a=1}
\frac{dx_\a}{2\pi ix_\a}
\frac{
\pl^k_{\a,\b}
\Delta
\(\frac{x_\a}{x_\b}\frac{1}{d}\frac{t}{q};q,t
\)}
{\pl^k_{\a\neq\b}
\Delta\(\frac{x_\a}{x_\b};q,t\)
}
\prod^k_{\a=1}
\left[
\prod^{N_c}_{l=1}
\Delta\(c\frac{x_\a}{a_l}\(\frac{q}{t}\)^{R_{\rm f}};q,t\)
\prod^{N_F-N_c}_{j=1}
\Delta\(c \frac{b_j}{x_\a} \(\frac{q}{t}\)^{R_{\rm af}};q,t\)
\right]
\nn\\
\end{eqnarray}
where $k$ denotes total vortex number,  $\kappa_2 = \frac{(q;q)^2_\infty}{\theta(t, q)}$, $c$ and $d$ denote the fugacity parameters for global $U(1)_c$ and $U(1)_Z$ respectively, and we have rewritten the fugacity parameter for the left-moving R-symmetry as $y=q^{1/2}/t$.
We have also used the function $\Delta(x;q,t)$ defined in (\ref{def-Delta}), where the first ratio corresponds to contributions from 
$\cN=(2,2)$ vector multiplet and adjoint chiral multiplet $Z^\alpha_\beta$, and the remainder come from the fundamental and anti-fundamental $\cN=(2,2)$ chiral multiplets.
Notice that under the shift $x_\a \to q x_\a$, the integral is invariant up to an overall factor $(t)^{N_f-2N_c}$, which implies that integrand is only an elliptic function of $x_\alpha$ for $N_f=2N_c$, and signals the breaking down of $U(1)_r$ as generated by $J_L$ to discrete subgroup ${\mathbb Z}_{2N_c-N_f}$.
In fact, this symmetry breaking pattern is precisely inherited from the bulk 4d $\cN=2$ theory, and we also expect the parameters $R_f$ and $R_{\rm af}$ to take the non-anomalous values if the theory flows to infra-red fixed point when mixes with other global abelian symmetries.
When we consider the simple poles from the $\cN=(2,2)$ adjoint chiral multiplet, they can be classified by length $N_c$ partitions of $k$, i.e. $\{\hn_l\}$ such that $\sum^{N_c}_{l=1} \hn_l=k$, and we can split $\{x_\alpha\}$ into $N_c$ different sets:
\begin{equation}
x_{l,\ga_l}= a_l
\left[c\(\frac{q}{t}\)^{{R_{\rm f}}}\right]^{-1} \td^{\ga_l}
,\quad \td = d\(\frac{q}{t}\), \quad \gamma_l=0,1\dots \hn_l-1.
\end{equation}
When we evaluate the residues at each of these simple poles, each of them corresponds to a Higgs vacuum of the $\cN=(2,2)$ vortex theory.
The 2d $\cN=(2,2)$ vortex world volume theory now reduces into:
\begin{eqnarray}\label{N=(2,2)Index-Int}
\cI^{\cN={(2,2)}}&=&
\sum\limits_{\{\hn_l\}}
\pl^{N_c}_{l,k=1}
\pl_{\gamma_l=0}^{\hn_l-1}
\pl_{\gamma_k=0}^{\hn_k-1}
\frac{
\Delta\(\td^{\gamma_l-\gamma_k-1}
\frac{a_l}{a_k};q,t\)}
{\Delta\(\td^{\gamma_l-\gamma_k}
\frac{a_l}{a_k};q,t\)}
\pl_{l,k=1}^{N_c}
\pl_{\gamma_l=0}^{\hn_l-1}
\Delta\(\td^{\gamma_l}
\frac{a_l}{a_k};q,t\)
\pl_{l=1}^{N_c}
\pl_{j=1}^{N_F-N_c}
\pl_{\gamma_l=0}^{\hn_l-1}
\Delta\(c^2\frac{b_j}{a_l}
\(\frac{q}{t}\)^{R_{\rm f}+R_{\rm af}}\td^{-\gamma_l};q,t\)
\nn\\
&=&
\sum_{\{\hn_l\}}
\pl^{N_c}_{l,k=1}
\pl^{\hn_l-1}_{\gamma_l=0}
\Delta
\(\frac{a_l}{a_k}
\td^{\gamma_l-\hn_k}
;q,t\)
\pl_{l=1}^{N_c}
\pl_{j=1}^{N_F-N_c}
\pl_{\gamma_l=0}^{\hn_l-1}
\Delta\(c^2\frac{b_j}{a_l}
\(\frac{q}{t}\)^{R_{\rm f}+R_{\rm af}}
\td^{-\gamma_l};q,t\).
\end{eqnarray}
Notice that in our calculation we have also introduced an additional global symmetry $U(1)_c$ as discussed earlier, 
which can be generated from appropriate linear combinations of $h_{34}$, $\fr$ and $\fR$.
However, its fugacity parameter $c$ merely appears as a spectator in the elliptic genus computation, 
and we can absorb it by overall shift of flavor fugacity parameters.
\paragraph{}
As a direct extension, we can also compute the elliptic genus for $\cN=(0,2)$ vortex theory, which can be viewed as a descendant of the $\cN=(2,2)$ vortex theory as 4d theories discussed earlier. The breaking of supersymmetry is done by turning on the superpotential for adjoint scalar in $\cN=(2,2)$ vector multiplet. 
Recall that, in 4d theories, we incorporate the effect of SUSY breaking superpotential  by setting $\ft=(\fp\fq)^{1-\frac{\tilde{\fr}_A}{2}}$, 
and the transformation properties of the matter fields under the residual R-symmetry are related by a single parameter $\tfr_A$.
This inspires us to consider the resultant $\cN=(0,2)$ vortex theory whose matter contents have global symmetry transformation properties
related by one parameter $\lambda$ as listed in Table \ref{table:N=(0,2)}. 
In particular, notice that $\lambda$ enters the charge assignments for both  $U(1)_Z$ and $U(1)_R$,
respectively labeled by fugacity parameters $d$ and $q^{1/2}$, indicating that we are only having one independent global $U(1)$ symmetry. 
\begin{table}[ht]
\centering
\begin{tabular}{|l|c|c|c|c|c|c|c|}
\hline  
& $Q$  &$\Lambda$& $\tilde{Q}$ &$\tilde{\Lambda}$&$\varphi$&$\varphi_\Lambda$&$\phi$\\\hline  
$U(k)_G$ &$\rm\bf k$&$\rm\bf k$ &$\bar{\rm\bf k}$ &$\bar{\rm\bf k}$&$\rm\bf adj.$&$\rm\bf adj.$&$\rm\bf adj.$\\\hline
$SU(N_c)_A$  & ${\rm\overline{\bf {N}}_c}$    & ${\rm\overline{\bf {N}}_c}$   & {\bf 1}& {\bf 1}& {\bf 1}& {\bf 1}& {\bf 1} \\\hline
$SU(N_F-N_c)_B $ & {\bf 1} & {\bf 1}  & ${\rm\bf {N_F-N_c} }$  & ${\rm\bf  {N_F-N_c} }$ & {\bf 1} & {\bf 1} & {\bf 1} \\\hline
$U(1)_Z$&$-h_{\rm f} \frac{\lam}{2}$&$(1-h_{\rm f}) \frac{\lam}{2}$&$-h_{\rm af} \frac{\lam}{2}$&$(1-h_{\rm af}) \frac{\lam}{2}$&$\frac{\lam}{2}-1$&$\lam-1$&$\frac{\lam}{2}$\\\hline
$U(1)_R$&$2h_{\rm f}(1-\lam)$&$R_{\rm Q}+2\lam-1$
&$2h_{\rm af}(1-\lam)$&$R_{\rm \tQ}+2\lam-1$&$2(\lam-1)$
&$4\lambda-3$&$2\lam$\\\hline
\end{tabular}
\caption{Field contents of $\cN=(0,2)$ vortex world volume theory, $R_{\rm Q}= 2h_{\rm f}(1-\lam)$ and $R_{\rm \tQ} = 2h_{\rm af}(1-\lam)$}
\label{table:N=(0,2)}
\end{table}
The elliptic genus for the $\cN=(0,2)$ vortex theory with these field contents can be written down as:
\begin{eqnarray}\label{N=(0,2)Index}
\cI^{\cN=(0,2)}_{\rm E}=
\frac{(q,q)^{2k}_\infty}{k!}
\oint&
\pl^k_{\a=1}&
\frac{dx_\a}{2\pi ix_\a}
\frac{
\pl^k_{\a\neq\b}\theta\(\frac{x_\a}{x_\b};q\)}{\pl^k_{\a\neq\b}
\theta\([d^{\frac{\lam}{2}}q^\lam]\frac{x_\a}{x_\b};q\)}
\pl^k_{\a,\b=1}
\frac{
\theta\(\frac{1}{dq}[d^{\frac{\lam}{2}}q^\lam]^{2}
\frac{x_\a}{x_\b};q\)}
{\theta\(\frac{1}{dq}[d^{\frac{\lam}{2}}
q^\lam]
\frac{x_\a}{x_\b};q\)}\nn
\\
\times
&\pl^k_{\a=1}&
\[
\pl^{N_c}_{l=1}
\frac{\theta\( \frac{x_\a}{ a_l}
d^{\frac{-h_{\rm f}\lambda}{2}} 
q^{h_{\rm f}(1-\lambda)}[d^{\frac{\lam}{2}}q^\lam];q\)}
{\theta\( \frac{x_\a}{ a_l}
d^{\frac{-h_{\rm f}\lambda}{2}}
q^{h_{\rm f}(1-\lambda)};q\)}
\pl^{N_F-N_c}_{j=1}
\frac{
\theta\(\frac{b_j}{x_\a} 
d^{\frac{-h_{\rm af}\lambda}{2}}
q^{h_{\rm af}(1-\lambda)}[d^{\frac{\lam}{2}}q^\lam];q\)}
{\theta\(\frac{b_j}{x_\a} 
d^{\frac{-h_{\rm af}\lambda}{2}}
q^{h_{\rm af}(1-\lambda)};q\)}\].\nn\\
\end{eqnarray}
Again if we consider the shift $x_\a \to q x_\a$, we notice that the integrand above picks up an overall factor $(d^{\frac{\lam}{2}}q^{\lam})^{N_f-2N_c}$.
This again implies the breaking of certain linear combination of global $U(1)$ symmetries with fugacity parameter $d^{\frac{\lam}{2}}q^\lam$ down to discrete ${\mathbb Z}_{N_f-2N_c}$.
However, if the theory flows to a superconformal fixed point, there needs to be an non-anomalous $U(1)$ R-symmetry completing the $\cN=(0,2)$ superconformal algebra.
We can again perform the contour integration as in the $\cN=(2,2)$ case, where the simple poles are classified by length $N_c$ partitions of $k$ i.e. $\{\hn_l\}$ such that
$\sum^{N_c}_{l=1} \hn_i=k$:
\begin{equation}
x_{l,\ga_l}=a_l d^{\frac{h_{\rm f}\lambda}{2}}q^{h_{\rm f}(\lambda-1)}
\td^{\gamma_i},
\quad 
\td=d\frac{q} {d^{\frac{\lam}{2}}q^\lam},
\quad
\gamma_l=0,1\cdots \hn_l-1.
\end{equation}
Summing up all the residues, the corresponding 2d $\cN=(0,2)$ elliptic genus reduces to:
\begin{eqnarray}\label{N=(0,2)Index-Int}
\cI^{\cN=(0,2)}_{\rm E}&=&
\sum\limits_{\{\hn_l\}}
\pl^{N_c}_{l,k=1}
\pl_{\ga_l=0}^{\hn_l-1}
\pl_{\ga_k=0}^{\hn_k-1}
\frac{\theta\(\td^{\ga_l-\ga_k}\frac{a_l}{a_k};q\)}
       {\theta\(\td^{\ga_l-\ga_k}\frac{a_l}{a_k}[d^{\frac{\lam}{2}}q^{\lambda}];q\)}
       \times
\pl^{N_c}_{l,k=1}
\pl_{\ga_l=0}^{\hn_l-1}
\pl_{\ga_k=0}^{\hn_k-1}
\frac{\theta\(\td^{\ga_l-\ga_k-1}\frac{a_l}{a_k}
[d^{\frac{\lam}{2}}q^{\lambda}];q\)}
       {\theta\(\td^{\ga_l-\ga_k-1}\frac{a_l}{a_k};q\)}
\nn\\
&&
\times
\pl_{l,k=1}^{N_c}
\pl_{\ga_l=0}^{\hn_l-1}
\frac{\theta\(\td^{\ga_l}
[d^{\frac{\lam}{2}}q^{\lambda}]
\frac{a_l}{a_k};q\)
}
{\theta\(\td^{\ga_l}\frac{a_l}{a_k};q\)}
\times
\pl_{l=1}^{N_c}
\pl_{j=1}^{N_F-N_c}
\pl_{\ga_l=0}^{\hn_l-1}
\frac{\theta\(\frac{b_j}{a_l} \td^{-\ga_l} d^{-\frac{(h_{\rm f}+h_{\rm af})\lambda}{2}}q^{(h_{\rm f}+h_{\rm af})(1-\lambda)}[d^{\frac{\lam}{2}}q^\lam];q\)}
       {\theta\(\frac{b_j}{a_l}\td^{-\ga_l}
        d^{-\frac{(h_{\rm f}+h_{\rm af})\lambda}{2}}q^{(h_{\rm f}+h_{\rm af})(1-\lambda)};q\)}
      \nn\\
&=&
\sum_{\{\hn_l\}}
\pl^{N_c}_{l,k=1}
\pl^{\hn_l-1}_{\ga_l=0}
\frac{\theta\(
\td^{\ga_l-\hn_k}
[d^{\frac{\lam}{2}}q^\lam]
\frac{a_l}{a_k};q\)}
{\theta\(\td^{\ga_l-\hn_k}\frac{a_l}{a_k};q\)}
\pl_{l=1}^{N_c}
\pl_{j=1}^{N_F-N_c}\pl_{\ga_l=0}^{\hn_l-1}
\frac{\theta\(\frac{b_j}{a_l} \td^{-\ga_l} d^{-\frac{(h_{\rm f}+h_{\rm af})\lambda}{2}}q^{(h_{\rm f}+h_{\rm af})(1-\lambda)}[d^{\frac{\lam}{2}} q^\lam];q\)}
       {\theta\(\frac{b_j}{a_l}\td^{-\ga_l}
        d^{-\frac{(h_{\rm f}+h_{\rm af})\lambda}{2}}q^{(h_{\rm f}+h_{\rm af})(1-\lambda)};q\)}.
        \nn\\
\end{eqnarray}
Interestingly, we notice that as its 4d counterparts, we can recover (\ref{N=(0,2)Index-Int}) from (\ref{N=(2,2)Index-Int}) by setting $t=d^{\frac{\lam}{2}}q^{\lambda}$, 
also setting $(R_{\rm f}; R_{\rm af})=(h_{\rm f}; h_{\rm af})$. In the next section, we will also compare this expression with the residues obtained from the $\cN=1$ superconformal index \eqref{4DN=1Index-Int}. 
\paragraph{}
Moreover we would like to compute the elliptic genus for the vortex world sheet theory arising from the 4d $\cN=1$ superconformal index of the 
magnetic dual theory,  under Seiberg/Kutasov-Schwimmer dualities (\ref{4DN=1Index-Int3}).
First, we note that the additional gauge singlet meson chiral fields are merely spectators which do not enter the contour integration, 
which allow us to recycle the calculation earlier and exchange the rank $N_c \to \tilde{N}_c = KN_f-N_c$.
Now similar discussion tells us that the resultant 2d vortex theory obtained this way would possess flavor symmetry group $SU(\tilde{N}_c) \times SU(N_f-\tilde{N}_c)$,
which sets constraint: $N_f - \tilde{N}_c\ge 0$ or $N_c \ge (K-1)N_f$.
Moreover, we also need $N_f \ge N_c$ for the vortex theory to exist in the electric theory. As a result, we conclude that we need $K=1$ for vortices to exist in both electric and magnetic theories, related by Seiberg duality \footnote{Another way to view this constraint is for the Higgs branches to exist in both electric and magnetic theories, which are needed for the dynamical vortices to exist.}.
For this case $\tilde{N}_c = N_f-N_c$, the corresponding elliptic genus can be written down analogously as:
\begin{eqnarray}
\cI^{\cN=(0,2)}_{\rm M}=
\frac{(q,q)^{2k}_\infty}{k!}
\oint&
\pl^k_{\a=1}&
\frac{dx_\a}{2\pi ix_\a}
\frac{
\prod^k_{\a\neq\b}\theta\(\frac{x_\a}{x_\b};q\)}{\pl^k_{\a\neq\b}
\theta\([d^{\frac{\lam}{2}}q^{\lam}]\frac{x_\a}{x_\b};q\)}
\prod^k_{\a,\b=1}
\frac{
\theta\(\frac{1}{dq}[d^{\frac{\lam}{2}}q^{\lambda}]^2
\frac{x_\a}{x_\b};q\)}
{\theta\(\frac{1}{dq}[d^{\frac{\lambda}{2}}
q^{\lambda}]
\frac{x_\a}{x_\b};q\)}
\nn\\
\times
&\pl^k_{\a=1}&\left[
\prod^{N_f-N_c}_{j=1}
\frac{\theta\( \frac{x_\a}{\tb_j} 
d^{\frac{-h_{\rm f}'\lambda}{2}} 
q^{h_{\rm f}'(1-\lambda)}[d^{\frac{\lambda}{2}}q^\lambda];q\)}
{\theta\( \frac{x_\a}{\tb_j}
d^{\frac{-h_{\rm f}'\lambda}{2}}
q^{h_{\rm f}'(1-\lambda)};q\)}
\prod^{N_c}_{l=1}
\frac{
\theta\(\frac{\ta_l}{x_\a } 
d^{\frac{-h_{\rm af}'\lambda}{2}}
q^{h_{\rm af}'(1-\lambda)}[d^{\frac{\lam}{2}}q^\lam];q\)}
{\theta\(\frac{\ta_l}{ x_\a }
d^{\frac{-h_{\rm af}'\lambda}{2}}
q^{h_{\rm af}'(1-\lambda)};q\)}\right].\nn\\
\end{eqnarray}
Similar calculation as before yields the following summation over the residues:
\begin{eqnarray}
\label{N=(0,2)Index-tilde}
{\cI}^{\cN=(0,2)}_{\rm M}&=&
\sum_{\{\hn_j'\}}
\pl^{N_f-N_c}_{i,j=1}
\pl^{\hn_i'-1}_{\ga_i=0}
\frac{\theta\(
\td^{\ga_i-\hn_j'}
[d^{\frac{\lambda}{2}}q^{\lambda}]
\frac{\tb_i}{\tb_j};q\)}
{\theta\(\td^{\ga_i-\hn_j'}\frac{\tb_i}{\tb_j};q\)}
\pl_{l=1}^{N_c}
\pl_{j=1}^{N_F-N_c}\pl_{\ga_j=0}^{\hn_j'-1}
\frac{\theta\(\frac{\ta_l} {\tb_j}\td^{-\ga_l} d^{-\frac{(h_{\rm f}'+h_{\rm af}')\lambda}{2}}q^{(h_{\rm f}'+h_{\rm af}')(1-\lambda)}[d^{\frac{\lam}{2}}q^\lam];q\)}
       {\theta\(\frac{\ta_l}{\tb_j}\td^{-\ga_l}
        d^{-\frac{(h_{\rm f}'+h_{\rm af}')\lambda}{2}}q^{(h_{\rm f}'+h_{\rm af}')(1-\lambda)};q\)}.
        \nn\\
\end{eqnarray}
The vacua are now classified by length $N_F-N_c$ partitions of $k$ i.e. $\{\hn_j'\}$ s.t. $\sum^{N_F-N_c}_{j=1} \hn_i'=k$.

\section{Matching 2d and 4d Indices and New 2d Dualities}\label{2d4dMatching}
\paragraph{}
From the calculations in the last two sections, we found that the 2d and 4d superconformal indices share almost identical functional forms.
In this section, we match them precisely by identifying the 2d and 4d fugacity parameters through their superconformal algebras. 
We will also rewrite the integrands in 4d superconformal indices into exponential form, and demonstrate that it can be factorized into two copies of four dimensional 
twisted superpotential on $T^2\times R^2$ considered in \cite{Nekrasov:2009uh} (See also Appendix \ref{Appendix-Factorization}). The natural fixed points in the localization computation correspond to surface operators wrapping on two intersecting $T^2$s.
We then consider the elliptic genera for the $\cN=(0,2)$ vortex theories in a pair of 4d $\cN=1$ supersymmetric theories connected by the Seiberg duality, 
and verify that they are related through a new $\cN=(0,2)$ duality, similar to non-abelian ``Hopping transition''. 
For comparison, we also present directly the Seiberg like duality directly on the 2d $\cN=(0,2)$ world volume theory.
\subsection{Embedding 2d Superconformal Algebras into 4d}
\begin{figure}[]
\centering
\includegraphics[scale=0.22]{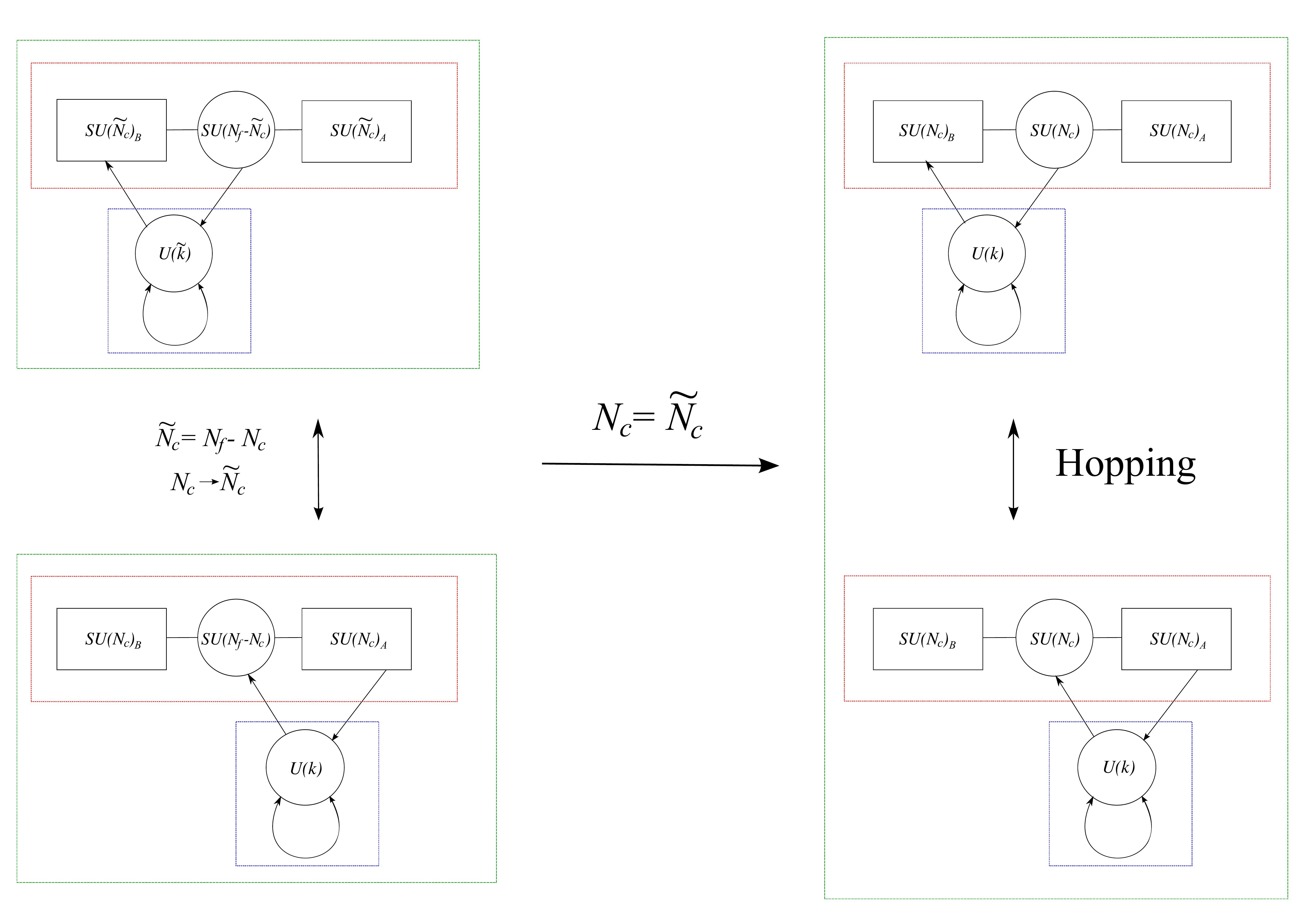}
\caption{Hopping of $\cN=(0,2)$ dynamical vortices/surface operators in electric/magnetic theories.}
\label{Hopping}
\end{figure}

\paragraph{}
Let us begin our discussion by recalling some basic facts about superconformal algebras used in computing 4d superconformal indices and 2d elliptic genera following \cite{Gadde:2010en, Gadde:2011uv, Gadde:2013dda}, and their connections.
For $\cN=2$ superconformal index, it counts the states which have vanishing value for one of the following eight combinations of conserved quantities:
\begin{eqnarray}\label{N=2Projections}
&&\delta_{1\pm} = E \pm (h_{12}-h_{34}) - 2\fR-\fr, \quad \delta_{2\pm} = E\pm (h_{12}-h_{34}) + 2\fR-\fr,\nn\\
&&\tilde{\delta}_{1\dot{\pm}} = E \pm (h_{12}+h_{34}) - 2\fR+\fr, \quad \tilde{\delta}_{2\dot{\pm}} = E\pm (h_{12}+h_{34})+ 2\fR+\fr,
\end{eqnarray} 
where we have used $2j_1=h_{12}-h_{34}$ and $2j_2=h_{12}+h_{34}$. Choosing a projection condition from above is equivalent to choosing a supercharge used in localization computation. 
Similarly for $\cN=1$ superconformal index, we have the the following four possible choices of projection conditions:
\begin{eqnarray}\label{N=1Projections}
\delta_{{\rm L} \pm} = E\pm (h_{12}-h_{34}) + \frac{3}{2} \tfr, \quad \delta_{{\rm R}\pm}=E\pm (h_{12}+h_{34}) - \frac{3}{2} \tfr.
\end{eqnarray} 
Recall in section \ref{4d-Indices}, we started from selecting $\tilde{\delta}_{1\dotdiv}$ in \eqref{defdeltaN=2} to compute $\cN=2$ superconformal index and then broke supersymmetry down to $\cN=1$, for which consistency from the angular momentum requires us to choose $\delta_{{\rm R}-}$ from (\ref{N=1Projections}) \footnote{Note that we exchanged the left and right handiness from \cite{Gadde:2010en}, in order to have consistent notation between 2d and 4d.} and the subsequent identification of R-charges (\ref{RchargeMap}) at superconformal fixed point.
\paragraph{}
We next consider the four possible projection conditions in the computation of elliptic genus of $\cN=(2,2)$ vortex theory, which are given by:
\begin{eqnarray}\label{N=(2,2)Projections}
\delta^{\cN=(2,2)}_{{\rm R}\pm} = 2H_{\rm R} \pm J_{\rm R}, \quad \delta^{\cN=(2,2)}_{{\rm L}\pm}= 2H_L \pm J_{\rm L}.
\end{eqnarray}
Finally for the elliptic genus of $\cN=(0,2)$ vortex theory, there are two possible projection conditions:
\begin{equation}\label{N=(0,2)Projections}
\delta^{\cN=(0, 2)}_{{\rm R}\pm} = 2H_{\rm R} \pm J_{\rm R}.
\end{equation}
As in the 4d cases, we need to consistently pick the projection conditions (\ref{N=(2,2)Cond}) and (\ref{N=(0,2)Cond}) while the theories involved are again related by superpotential deformation.
In order to match the 4d/2d fugacity parameters with respect to these projection conditions, there are two equivalent ways to identify the conserved quantities in 4d $\cN=2$ and 2d $\cN=(2,2)$ theories following \cite{Gadde:2013dda}, i. e.
\begin{eqnarray}
&&E-h_{12}=2 H_{\rm R}, \quad 2\fR-\fr+h_{34} = J_{\rm R},\label{matchCond1} \\
&&E-h_{34}=2 H_{\rm R}, \quad 2\fR-\fr+h_{12} = J_{\rm R}.\label{matchCond2}
\end{eqnarray}
They precisely correspond to the two complementary ways to insert co-dimension two defects into two orthogonal planes $x^{1,2}$ and $x^{3,4}$.
Let us consider first the surface operator inserted in $x^{1,2}$ plane with identification given by (\ref{matchCond1}), where $h_{12}$ now descends as the angular momentum in the 2d world sheet of the surface defect labeled by $q$ while the transverse angular momentum $h_{34}$ now mixes with $(\fR, \fr)$ to give the 2d $U(1)_{\rm R}$ generator $J_{\rm R}$.
{Finally, from parameter counting, we also need  another 2d fugacity parameter $d$ to be generated by the linear combination $\fR+h_{34}$ and denote this symmetry as $U(1)_z$.}
Simple algebra then shows that modulo additional flavor symmetries, we can identify the 2d and 4d fugacity parameters as:
\begin{equation}\label{N=2Matching}
q=\fq,\quad d=\frac{\fp^2}{\ft}, \quad t=\frac{\fp\fq}{\ft}.
\end{equation}
Now if we compare the expressions in (\ref{EGN=2}) and (\ref{N=(2,2)Index-Int}), we can identify them precisely by setting \footnote{We have set the fugacity for $U(1)_c$ symmetry $c$ to 1. }:
\begin{equation}\label{N=2Matching2}
 n_{\alpha} = \hat{n}_l, \quad \fa_{\alpha} = a_l, \quad \fa_j \fp^{-1}= b_j\(\frac{q}{t}\)^{R_{\rm f}+R_{\rm af}}   
\end{equation}
and noticing that $\alpha \in \{i_l\}$ and $j \in \overline{\{i_l\}}$, $l=1,\dots, N_c$. We can make similar identification for the other set of surface operators inserted in $x^{3,4}$ plane by exchanging $\fp \leftrightarrow \fq$.
Consequently, we have thus successfully confirmed that for arbitrary $(N_c, N_f)$, the theta-function dependent parts of the residues of $\cN=2$ superconformal indices reduce to two copies of the elliptic genus of the $\cN=(2,2)$ vortex world sheet theory.
\paragraph{}
We can also perform analogous analysis to match the fugacity parameters in 4d $\cN=1$ superconformal index and 2d $\cN=(0,2)$ elliptic genus,
where we need additional 2d fugacity parameter $e$ which is generated by $h_{34}+\frac{\tfr}{2}$ from the parameter counting. As a result, we have:
\begin{equation}\label{N=1Matching}
q=\fq, \quad e = \frac{\fp}{\fq^{1/2}}.
\end{equation}
Now if we recall that we implemented the effect of 4d supersymmetry breaking superpotential by setting $\ft=(\fp\fq)^{1-\frac{\tfr_{\rm A}}{2}}$, 
\eqref{N=2Matching} tell us the analogous 2d effect is to setting $t=d^{\frac{\lambda}{2}}q^\lambda$ as noted earlier.
Furthermore, to match the residues in one of the last two lines of integrated $\cN=1$ superconformal index (\ref{4DN=1Index-Int}) with the $\cN=(0,2)$ elliptic genus (\ref{N=(0,2)Index-Int}), 
we need to further identify the parameters:
\begin{equation}\label{N=1Matching2}
{\lambda} = \frac{{\tfr_{\rm A}}}{1+\frac{\tfr_{\rm A}}{2}},
\end{equation}
along with the following\footnote{Here we have also related the fugacity parameters as $e=d^{1-\frac{\lam}{2}} q^{-\lam}$.}:
\begin{equation}
 n_{\alpha} = \hat{n}_l, \quad \fa_{\alpha} = a_l, \quad \fa_j \fp^{-1}= b_j d^{-\frac{(h_{\rm f}+h_{\rm af})\lam}{2}} q^{(h_{\rm f}+h_{\rm af})(1-\lam)}
\end{equation}
where  $\alpha \in \{i_l\}$ and $j \in \overline{\{i_l\}}$, $l=1,\dots, N_c$.
Again we have shown that for arbitrary $(N_f, N_c)$, provided the R-charge constraint from superpotential (\ref{RChargeCond}) is satisfied, 
the theta function dependent parts of the residue of $\cN=1$ superconformal index coincide with two copies of the elliptic genus of the $\cN=(0,2)$ vortex theory.   
Alternatively, if we set $K=1$ and $\frac{\tfr_{\rm A}}{2} = 1-\frac{(\tfr_{\rm f}'+\tfr_{\rm af}')}{2}$ from the $\cN=2$ superpotential constraint in the $\cN=1$ superconformal index of the magnetic dual theory (\ref{4DN=1Index-Int3}), we can also match it with \eqref{N=(0,2)Index-tilde} by the following identification of 2d/4d flavor fugacity parameters:
\begin{equation}
n_j' = \hat{n}_j' ,\quad  \fa_j \fp^{-1} = \tb_j^{-1} d^{-\frac{(h_{\rm f}+h_{\rm af})\lam}{2}} q^{(h_{\rm f}+h_{\rm af})(1-\lam)}, \quad  \fa_\a = \ta_l^{-1}, \quad 
\end{equation}
where $j \in \{i_k\}, \a \in \overline{\{i_k\}}$ and $k=1,\dots N_f-N_c$ \footnote{Here we have swapped the indices $\a$ and $j$ from earlier 4d index in \eqref{4DN=1Index-Int3} to make the comparison with the electric theory more transparent.}.
Now we clearly see that the elliptic genera for the vortices in electric/magnetic theories are again related by Seiberg duality acting on their fugacity parameters $\{a_l; b_j\} \leftrightarrow \{1/\ta_l; 1/\tb_j  \}$ and $N_c \leftrightarrow N_f-N_c$. But we can also regard such transformation as two different ways to gauge the flavor symmetries of the same 2d $\cN=(0,2)$ vortex theory and embed it into two different 4d $\cN=1$ SQCDs as illustrated in the left side of Figure \ref{Hopping}.
Let us now recall in checking 4d $\cN=1$ Seiberg duality, we need to assign non-anomalous $U(1)_R$ charges to the matter fields for the electric and magnetic superconformal indices to match, where these two constraint can be satisfied simultaneously if $N_f = 2N_c$ or $\tilde{N}_c = N_c$. In such a limit, the elliptic genera for the $\cN=(0,2)$ vortex theories in electric/magnetic theories, \eqref{N=(0,2)Index-Int} and \eqref{N=(0,2)Index-tilde}, are mapped into each other under $\{a_l; b_j\} \leftrightarrow \{1/\ta_l; 1/\tb_j  \}$ transformation, and we have seen that 4d Seiberg duality now descend down to the $\cN=(0,2)$ vortex theory as the invariance of its elliptic genus under ``hopping'' along the quiver nodes, as illustrated on the right side of Figure \ref{Hopping}. We can also regard this as a statement that the ``self-adjointness'' of the difference operator, corresponding to the insertion of vortices into 4d $\cN=2$ superconformal index, still holds under the breaking down to $\cN=1$.
\paragraph{}
We can also understand the structure of the 4d superconformal index and the matching with 2d elliptic genus by rewriting the integrand in (\ref{4DN=2Index}) into manifestly factorizable form.
While relegating most of the calculation details in a self-contained appendix \ref{Appendix-Factorization}, we present the final result here:
\begin{eqnarray}
&&\cI^{\cN=2} =\frac{\kappa_4^{N_c}}{N_c !} \int \prod_{l=1}^{N_c} d\sigma_l \exp\(-i\frac{\pi}{3}\cF[\{\sigma_l\}, \{m_i\}] \)\nn\\
&&\times
\pl^{N_c}_{l,k=1}
\pl_{s\in {\mathbb Z}}
\exp\(\sum^\infty_{m=0}\frac{B_m(\log \rq)^{m-1}}{m!}
\[
-
{\rm Li_{2-m}}\(e^{2\sqrt{2}\pi b {\D}}\)
+
{\rm Li_{2-m}}\(e^{2\pi b \(\s_k-\s_l+\frac{s}{\e}-2\D\)}\rq\)
-
{\rm Li_{2-m}}\(e^{2\pi b\( \s_k-\s_l+\frac{s}{\e}\)}\rq\)
\]\)\nn\\
&&\times
\pl^{N_f}_{i=1}
\pl^{N_c}_{l=1}
\pl_{s\in{\mathbb Z}}
\exp\(
\sum^\infty_{m=0}
\frac{B_m(\log \rq)^{m-1}}{m!}
\[
{\rm Li_{2-m}}\(e^{2\pi b
(-\s_l+m_i+\D)}\rq\)
-
{\rm Li_{2-m}}\(e^{2\pi b
(-\s_l+m_i)}\rq\)
-
{\rm Li_{2-m}}\(e^{2\pi b
{\D}}\rq\)
\right.\right.
\nn\\
&&
\quad\qquad\qquad\qquad\qquad\qquad\qquad
\left.\left.
\left.
+
{\rm Li_{2-m}}\(e^{2\pi b\(
-\s_l+m_i+i(b^{-1}-J)+
\frac{s}{\e}\)}\rq\)
-
{\rm Li_{2-m}}\(e^{2\pi b\(
-\s_l+m_i+i(b^{-1}-J)+
\frac{s}{\e}-2\D\)}\rq\)
\]\)
\right\}
\nn\\
&&
\times \{b\rightarrow b^{-1}\}.
\end{eqnarray}
In above, we have defined the following parameters:
\begin{eqnarray}\label{FactorizationParameters}
&&
{\rm q} = e^{2\pi i b^2},\quad
z_l=e^{2\pi i \sigma_l},\quad 
\fa_i=e^{2\pi i m_i},
\quad \ep=\frac{\tilde{r}_1}{r_3},
\nn\\
&&\fp = e^{-2\pi \ep b}, \quad \fq=e^{-2\pi \ep/b},\quad
\sqrt{\ft}=e^{-2\pi\epsilon  J}, \quad \Delta =\frac{1}{4\pi i\epsilon} \log\frac{\fp\fq}{\ft},
\end{eqnarray}
and the phase factor $\cF [\{\sigma_l\}, \{m_i\}]$ is defined to be
\begin{equation}
\mathcal{F}[\{\sigma_l\}, \{m_i\}] = \sum_{l, k=1}^{N_c} (B_{33}[\ep(\s_l-\s_k+2\Delta)]-B_{33}[\ep(\s_l-\s_k)])
-\sum_{l=1}^{N_c}\sum_{i=1}^{N_f} (B_{33}[\ep(\s_l-m_i+iJ+2\Delta)]-B_{33}[\ep(\s_l-m_i+iJ)])
\end{equation}
with the function $B_{33}[x] = B_{33}(x|1, i\e b, i\e/b)$ defined in (\ref{DefB33-2}).
Other than the phase factor $\cF(\{\s_l\};\{m_i\})$, we have factorized the integrand into purely $b$ and $1/b$ dependent pieces under the changes of variables \eqref{FactorizationParameters}. 
It is explained in more details in Appendix \ref{Appendix-Factorization} that due to the presence of background fluxes which deform $S^3$ into three dimensional ellipsoid $S_b^3$, 
we can regard the 4d superconformal indices as the partition functions on $\tilde{S}^1 \times S_b^3$, where $\tilde{S}^1$ is a trivially fibered circle over $S^3_b$, whose radius $\tilde{r}_1$ depends on the 3d deformation parameter $b$ as defined in \eqref{3d4dconversion2}. 
Now we recall from \cite{Pasquetti:2011fj, Beem:2012mb} that the partition function on $S_b^3$ \cite{Hama:2011ea} can be factorized into two copies of twisted partition functions on $S^1\times_{\rm q} D^2$, glued together by the so-called ``S-fusion''  mapping $b \to 1/b$ or 
${\rm q} = e^{2i\pi b^2} \to \tilde{{\rm q}} = e^{2i\pi/b^2 }$. It has been explicitly shown in \cite{Chen:2013pha, Fujitsuka:2013fga, Benini:2013yva} that each copy of partition function on $S^1 \times_{\rm q} D^2$ can be reproduced by summing over BPS vortices (anti-vortices) wrapping along $S^1$, which is sometimes called ``Higgs branch'' localization as the fixed points now precisely correspond to these vortices on Higgs branch. Now adding another trivially fibered $\tilde{S}^1$, we are promoting the the world volumes of vortices into a pair of two tori, whose complex structure modulus can be given by either $\fp = e^{-2\pi\epsilon b}$ or $\fq = e^{-2\pi \epsilon/b}$, namely the surface operators which we have been dealing with. Note that the two different surface operators actually intersect on a common $\tilde{S}^1$, and we expect their interaction giving rise to the non-factorizable phase factor $\cF(\{\sigma_l\}; \{m_l\})$, as in the Abelian case where such a factor (proportional to $\frac{\fp\fq}{\ft}$) can be generated by Chern-Simons like interaction between ``linked vortices'' \cite{Gaiotto:2012xa}. It would be interesting to understand the origin of this phase factor in the non-Abelian theories better, as it is responsible for the non-factorizable factors \eqref{nonfactor1} encountered in the earlier section.

\subsection{Hori-Tong Duality for $\cN=(0,2)$ Theory}
\paragraph{}
Here we also consider the Seiberg-like duality in two dimensions, focusing on the theories with two supercharges, i.e 2d $\cN=(0,2)$ theories and their elliptic genera. 
For original Seiberg-like duality in $\cN=(2,2)$ theories (sometimes also known as Hori-Tong Duality \cite{Hori:2006dk}), there are generally singlet meson fields appear in the dual magnetic theories.
However, here as in the 2d vortex theories discussed earlier, we introduce polynomial superpotential for the $\cN=(2,2)$ vector multiplets in these theories and break supersymmetry explicitly down to $\cN=(0,2)$. 
We will see that similar duality still holds, and there are additional  "meson Fermi multiplets" appearing in the resultant dual magnetic theory.
Starting from the electric theory whose matter contents and their symmetry properties are listed in Table \ref{Table: N=(0,2) H-T electric}, denoting the fugacity for flavor symmetries $SU(N)_A\times SU(N)_B\times U(1)_c$ as $(a_l,b_l,c)$, we can write down the index:
\begin{table}[ht]
\centering
\begin{tabular}{|l|c|c|c|c|c|}
\hline  
& $Q$  &$Q_{\Lambda}$& $\tilde{Q}$ &$\tilde{Q}_{\Lambda}$&$\phi$\\\hline  
$U(M)_G$ &$\rm\bf M$&$\rm\bf M$ &$\overline{\rm\bf M}$ &$\overline{\rm\bf M}$&$\rm\bf adj.$\\\hline
$SU(N)_A$  & ${\rm\bf {N}}$    & ${\rm\bf {N}}$   & {\bf 1}& {\bf 1}& {\bf 1}\\\hline
$SU(N)_B $ & {\bf 1} & {\bf 1}  & ${\rm\bf \overline{N} }$  & ${\rm\bf  \overline{N} }$ & {\bf 1} \\\hline
$U(1)_c$&1&1&1&1&0
\\\hline
$U(1)_Z$&$-h_{\rm f} \frac{\lam}{2}$&$(1-h_{\rm f}) \frac{\lam}{2}$&$-h_{\rm af} \frac{\lam}{2}$&$(1-h_{\rm af}) \frac{\lam}{2}$&$\frac{\lam}{2}$\\\hline
$U(1)_R$&$2h_{\rm f}(1-\lam)$&${\rm R_Q}+2\lam-1$
&$2h_{\rm af}(1-\lam)$&${\rm R_{\tQ}}+2\lam-1$&$2\lam$\\\hline
\end{tabular}
\caption{Field contents of $\cN=(0,2)$ electric theory, where $R_Q = 2h_{\rm f}(1-\lam)$ and $R_{\tQ} =2h_{\rm af}(1-\lam)$. }
\label{Table: N=(0,2) H-T electric}
\end{table}
\begin{eqnarray}
&&\cI_E^{\cN=(0,2)}=\cI_{k}(a,b,c;q)
\nn\\
&&
=\frac{(q,q)^{2M}_\infty}{M!}
\oint
\pl^M_{\a=1}
\frac{dx_\a}{2\pi ix_\a}
\frac{
\pl^M_{\a\neq\b}\theta\(\frac{x_\a}{x_\b};q\)}{\pl^M_{\a,\b=1}
\theta\(\frac{x_\a}{x_b}[d^{\frac{\lam}{2}}q^{\lam}];q\)}
\pl^M_{\a=1}
\pl^{N}_{l=1}
\frac{\theta\(c x_\a a_l
d^{\frac{-h_{\rm f}\lambda}{2}} 
q^{h_{\rm f}(1-\lambda)}[d^{\frac{\lambda}{2}}q^\lambda];q\)\theta\(\frac{c}{b_lx_\a} 
d^{\frac{-h_{\rm af}\lambda}{2}}
q^{h_{\rm af}(1-\lambda)}[d^{\frac{\lambda}{2}}q^\lambda];q\)}
{\theta\( c x_\a a_l
d^{\frac{-h_{\rm f}\lambda}{2}}
q^{h_{\rm f}(1-\lambda)};q\)\theta\(\frac{c}{b_l x_\a} 
d^{\frac{-h_{\rm af}\lambda}{2}}
q^{h_{\rm af}(1-\lambda)};q\)}
\nn\\
\end{eqnarray}
By taking the poles from the the fundamental chiral multiplets, we evaluate the residue at which the poles satisfy the following condition, by picking $M$ out of the $N$ possible choices: 
\begin{equation}
x_\a =\frac{1}{ca_{l_\a}} d^{\frac{h_{\rm f}\lambda}{2}}
q^{h_{\rm f}(\lambda-1)}, \quad \a=1,2\cdots M.
\end{equation}
Here we again use the notations of $\{l_\a\}$ and $\overline{\{l_\a\}}$ to denote the chosen part and its complement.
The contour integration now reduces to a summation of residues at simple poles:
\begin{eqnarray}
\label{(0,2) Hori-Tong electric}
\cI^{\cN=(0,2)}_{\rm Electric}&=&
\sum_{\{l_\a\}}
\pl^M_{\a\neq\b}
\frac{\theta\(\frac{a_{l_\b}}{a_{l_\a}};q\)}
       {\theta\(\frac{a_{l_\b}}{a_{l_\a}}
       [d^{\frac{\lam}{2}}q^{\lam}];q\)}
\pl^M_{\a =1}
\pl_{j \in \overline{\{l_\a\}} }
\frac{\theta\(\frac{a_j}{a_{l_\a}}[d^{\frac{\lam}{2}}q^{\lam}];q\)}{\theta\(\frac{a_j}{a_{l_\a}};q\)}
\pl^M_{\a=1}
\pl^N_{k=1}
\frac{\theta\(c^2\frac{a_{l_\a}}{b_k}
d^{\frac{-(h_{\rm f}+h_{\rm af})\lambda}{2}}
q^{(h_{\rm f}+h_{\rm af})(1-\lambda)}[d^{\frac{\lam}{2}}q^\lam];q\)}     
       {\theta\(c^2\frac{a_{l_\a}}{b_k}d^{\frac{-(h_{\rm f}+h_{\rm af})\lambda}{2}}
q^{(h_{\rm f}+h_{\rm af})(1-\lambda)};q\)}
\nn\\
&=&
\sum_{\{l_\a\}}
\pl_{s\in\{l_\a\}}
\pl^N_{k=1}
\frac{\theta\(c^2\frac{a_s}{b_k}
d^{\frac{-(h_{\rm f}+h_{\rm af})\lambda}{2}}
q^{(h_{\rm f}+h_{\rm af})(1-\lambda)}[d^{\frac{\lam}{2}}q^\lam];q\)}     
       {\theta\(c^2\frac{a_s}{b_k}d^{\frac{-(h_{\rm f}+h_{\rm af})\lambda}{2}}
q^{(h_{\rm f}+h_{\rm af})(1-\lambda)};q\)}
\pl_{s\in \{l_\a\}}
\pl_{r\in\overline{\{l_\a\}}}
\frac{\theta\(\frac{a_r}{a_s}[d^{\frac{\lam}{2}}q^{\lam}];q\)}{\theta\(\frac{a_r}{a_s};q\)}.
\end{eqnarray}
To make it more symmetrical to read off the duality, one can use a theta function identity $\theta(zq;q)=\theta(z^{-1};q)$ to rewrite the first factor above into  (for fixed subscript $k$ ):
\begin{eqnarray}
\label{meson trick}
&&\pl_{s\in\{l_\a\}}
\frac{\theta\(c^2\frac{a_s}{b_k}
d^{\frac{-(h_{\rm f}+h_{\rm af})\lambda}{2}}
q^{(h_{\rm f}+h_{\rm af})(1-\lambda)}[d^{\frac{\lam}{2}}q^\lam];q\)}     
       {\theta\(c^2\frac{a_s}{b_k}d^{\frac{-(h_{\rm f}+h_{\rm af})\lambda}{2}}
q^{(h_{\rm f}+h_{\rm af})(1-\lambda)};q\)}\nn\\
&&=
\pl_{l=1}^N
\frac{\theta\(c^2\frac{a_l}{b_k}
d^{\frac{-(h_{\rm f}+h_{\rm af})\lambda}{2}}
q^{(h_{\rm f}+h_{\rm af})(1-\lambda)}[d^{\frac{\lam}{2}}q^\lam];q\)}     
       {\theta\(c^2\frac{a_l}{b_k}d^{\frac{-(h_{\rm f}+h_{\rm af})\lambda}{2}}
q^{(h_{\rm f}+h_{\rm af})(1-\lambda)};q\)}
\pl_{r\in \overline{\{l_\a\}}}
\frac{\theta\(\frac{1}{c^2}\frac{b_k}{a_r}
d^{\frac{(h_{\rm f}+h_{\rm af})\lambda}{2}}
q^{(h_{\rm f}+h_{\rm af})(\lambda-1)+1};q\)} 
       {\theta\(\frac{1}{c^2}\frac{b_k}{a_r}
d^{\frac{(h_{\rm f}+h_{\rm af})\lambda}{2}}
q^{(h_{\rm f}+h_{\rm af})(\lambda-1)+1}[d^{-\frac{\lam}{2}}q^{-\lam}];q\)}.
\nn\\ 
\end{eqnarray}
Now we can consider the elliptic genus for the dual magnetic theory, also obtained from the superpotential deformation.
We first consider changing the rank of gauge group $N \rightarrow N-M$ to obtain:
\begin{eqnarray}
\cI_{N-M}(\ta,\tb,\tc;q)
=
\sum_{\overline{\{l_\a\}}}
\pl_{r\in\overline{\{l_\a\}}}
\pl^N_{l=1}
\frac{\theta\(\tc^2\frac{\ta_r}{\tb_l}
d^{\frac{-(h_{\rm f}+h_{\rm af})\lambda}{2}}
q^{(h_{\rm f}+h_{\rm af})(1-\lambda)}[d^{\frac{\lam}{2}}q^\lam];q\)}     
       {\theta\(\tc^2\frac{\ta_r}{\tb_l}
       d^{\frac{-(h_{\rm f}+h_{\rm af})\lambda}{2}}       
       q^{(h_{\rm f}+h_{\rm af})(1-\lambda)};q\)}
\pl_{r\in\overline{\{l_\a\}}}
\pl_{s\in \{l_\a\}}
\frac{\theta\(\frac{\ta_s}{\ta_r}
[d^{\frac{\lambda}{2}}
q^{\lambda}];q\)}
{\theta\(\frac{\ta_s}{\ta_r};q\)}.
\nn\\
\end{eqnarray}
Now by applying the identity \eqref{meson trick}, we can write down the relation for these two theories:
\begin{equation}
\label{(0,2) Hori-Tong dual}
\cI^{\cN=(0,2)}_{\rm Electric}=\cI_{M}(a,b,c;q)=
\(\pl_{l,k=1}^N
\frac{\theta\(\frac{1}{\tc^{2}}\frac{\tb_l}{\ta_k}
       d^{\frac{(h_{\rm f}+h_{\rm af})\lambda}{2}}
       q^{(h_{\rm f}+h_{\rm af})(\lambda-1)+1};q\)}
{\theta\(\frac{1}{\tc^2}
            \frac{\tb_l}{\ta_k}
            d^{\frac{(h_{\rm f}+h_{\rm af})\lambda}{2}}            
            q^{(h_{\rm f}+h_{\rm af})(\lambda-1)+1}[d^{-\frac{\lam}{2}}q^-\lam];q\)}
       \)
\cI_{N-M}(\ta,\tb,\tc;q)
=\cI^{\cN=(0,2)}_{\rm Magnetic}
\end{equation}
where $\ta_l:=1/a_l$ $\tb_l:=1/b_l$ $\tc^2:=c^{-2}d^{((h_{\rm f}+h_{\rm af})-\frac{1}{2})\lambda}q^{(2(h_{\rm f}+h_{\rm af})-1)(\lambda-1)}$. On the right hand side, the extra ratio now corresponds to the singlet chiral meson fields $\cM_{l}^k$ and "mesonic Fermi multiplets $\cM_{\Lambda,l}^k$", while the symmetry properties of the fields in this theory are now summarized in Table \ref{Table: N=(0,2) H-T magnetic}.
\begin{table}[ht]
\centering
\begin{tabular}{|l|c|c|c|c|c|c|c|}
\hline  
& $q$  &$q_{\Lambda}$& $\tilde{q}$ &$\tilde{q}_{\Lambda}$&$\phi$&$\cM$&$\cM_\Lambda$\\\hline  
$U(N-M)_{\tilde{G}}$ &$\rm\bf N-M$&$\rm\bf N-M$ &$\overline{\rm\bf N-M}$ &$\overline{\rm\bf N-M}$&$\rm\bf adj.$&0&0\\\hline
$SU(N)_A$  & ${\rm\bf {N}}$    & ${\rm\bf {N}}$   & {\bf 1}& {\bf 1}& {\bf 1}& ${\rm\bf \overline{N} }$  & ${\rm\bf  \overline{N} }$\\\hline
$SU(N)_B $ & {\bf 1} & {\bf 1}  & ${\rm\bf \overline{N} }$  & ${\rm\bf  \overline{N} }$ & {\bf 1}& ${\rm\bf {N}}$    & ${\rm\bf {N}}$  \\\hline
$U(1)_c$&1&1&1&1&0&-2&-2
\\\hline
$U(1)_Z$&$-\frac{h_{\rm f} \lam}{2}$&$\frac{(1-h_{\rm f})\lam}{2}$&$-\frac{h_{\rm af} \lam}{2}$&$\frac{(1-h_{\rm af}) \lam}{2}$&$\frac{\lam}{2}$&$\frac{(h_{\rm f}+h_{\rm af}-1)\lambda}{2}$ &$\frac{(h_{\rm f}+h_{\rm af})\lambda}{2}$\\\hline
$U(1)_R$&$2h_{\rm f}(1-\lam)$&${\rm R_Q}+2\lam-1$
&$2h_{\rm af}(1-\lam)$&${\rm R_{\tQ}}+2\lam-1$&$2\lam$
&$R_{\cM}$&$R_{\cM}+2\lambda-1$\\\hline
\end{tabular}
\caption{Field contents of $\cN=(0,2)$ magnetic theory, $ R_{\cM}= 2(h_{\rm f}+h_{\rm af}-1)(\lambda-1)$}
\label{Table: N=(0,2) H-T magnetic}
\end{table}
The duality relation \eqref{(0,2) Hori-Tong dual} tells us that the $\cN=(0,2)$  $U(M)$ gauge theory is dual to $U(N-M)$ gauge theory  with the same number $N$ of matter fields plus additional singlet meson fields. As only fermionic superpotential couplings are allowed in $\cN=(0,2)$ theory, we expect the existence of superpotential coupling of the form $\cW(\cM,q,\tilde{q}) \sim q_i\cM_{ij}\tilde{q}_{\Lambda,j}
      +q_i\cM_{\Lambda,ij}\tilde{q}_j
      +q_{\Lambda,i}\cM_{ij}\tilde{q}_j$ 
{and that the R-charge assignment above reflect this, as each term in the superpotential now summed up to 1.} Furthermore, such a $\cN=(0,2)$ superpotential  can also be deduced by decomposing the $\cN=(2,2)$ meson superpotential into $(0,2)$ components. We have thus generalized the Hori-Tong duality to the $\cN=(0,2)$ theories via superpotential deformation.

\subsection{Coupling $\cN=(0,2)$ Duality to 4d Theory}
\paragraph{}
In the previous sections, we have demonstrated how to obtain 2d surface operators which can be interpreted as the IR limit of dynamical vortices, from the residue of 4d superconformal indices. Moreover, from the residues for the quiver gauge theories, we can embed this class of surface defects to 4d gauge theory by gauging their flavor symmetries to become part of the 4d gauge group.  
However we can also apply the gauging procedure to couple other surface defects which do not correspond to IR limit of dynamical vortices, but rather appear as the boundary conditions for the bulk fields on two out of four dimensions, as done in \cite{Gadde:2013dda} for coupling 2d $\cN=(2,2)$ gauge theories to 4d $\cN=2$ Superconformal QCD (See \cite{Gomis:2014eya} for detailed discussion on recent developments.). 
In particular, 
this allows us to beautifully re-interpret 2d Hori-Tong duality involving $\cN=(2,2)$ SQCD as the invariance of the corresponding surface defects under the generalized S-duality transformation of bulk 4d $\cN=2$ superconformal field theories, and we can deduce a ``triality relation'' among three different $\cN=(2,2)$ theories.
{Here we would like to couple the theories involved in the generalized 2d $\cN=(0,2)$ Hori-Tong duality discussed in the last section, to the 4d $\cN=1$ gauge theory which can be reproduced by SUSY breaking superpotential from $\cN=2$ superoconformal QCD.} We will see that the aforementioned triality relation still survives under such superpotential deformation.
\paragraph{}
To do so, we need to first apply 2d/4d fugacity parameter mapping $q=\fq,~ d^{\frac{\lambda}{2}}q^\lambda=(\fp\fq)^{\frac{\tr_A}{2}}$ and to identify one of the 2d flavor fugacity $a_l$ with the parameter $z_l$, i. e.  $\cI_{\rm Electric}= \cI_M (a=z,b=\fb,c)$. Insetting this into $\cN=1$ superconformal index for the superpotential deformed theory \eqref{N=1IndexElectric}, we also need to change the gauge group from $U(N)$ to $SU(N)$, also set $N_f=2N$ and impose the R-charge constraint \eqref{RChargeCond}. {Without ambiguity and for simplicity, we drop $(\fp;\fq)$ in elliptic gamma function, denoting $\Gamma(z;\fp,\fq)=\Gamma(z)$ thus obtain}:
\begin{eqnarray}
\label{2d4d-electric}
\cI_{2d-4d}
&=&
\frac{\kappa^{N-1}}{N!}
\oint_{\bbT^{N-1}}
\pl^{N-1}_{l=1}
\frac{dz_l}{2\pi i z_l}
\frac{\pl_{l,k=1}^N\Gamma\((\fp\fq)^{\frac{\tr_A}{2}}\frac{z_l}{z_k}\)}
{\pl_{l\neq k =1}^N\Gamma\(\frac{z_l}{z_k}\)}
\pl_{l,k = 1}^{N}
\Gamma\(\frac{z_l}{\fa_k}
\(\fp\fq\)^{\frac{\tfr_{{\rm f}_a}}{2}}\)
\Gamma\(\frac{\fa_k}{z_l}
\(\fp\fq\)^{\frac{\tfr_{{\rm af}_a}}{2}}\)
\Gamma\(\frac{z_l}{\fb_k}
\(\fp\fq\)^{\frac{\tfr_{{\rm f}_b}}{2}}\)
\Gamma\(\frac{\fb_k}{z_l}
\(\fp\fq\)^{\frac{\tfr_{{\rm af}_b}}{2}}\)
\nn\\
&\times&
\sum_{\{l_\a\}}
\pl_{s\in\{l_\a\}}
\pl^N_{k=1}
\frac{\theta\(c^2\frac{z_s}{b_k}
\fq^{(h_f+h_{af})}
(\fp\fq)^{\frac{\tr_A\(1-(h_{\rm f}+h_{\rm af})\)}{2}};\fq\)}     
       {\theta\(c^2\frac{z_s}{b_k}
\fq^{(h_f+h_{af})}
(\fp\fq)^{\frac{-\tr_A(h_{\rm f}+h_{\rm af})}{2}} ;\fq\)}
\pl_{s\in \{l_\a\}}
\pl_{r\in\overline{\{l_a\}}}
\frac{\theta\(\frac{z_r}{z_s}(\fp\fq)^{\frac{\tr_A}{2}};\fq\)}{\theta\(\frac{z_r}{z_s};\fq\)}
\end{eqnarray}
where $\a=1,2,\cdots k$ and $\kappa= (\fp, \fq)_\infty (\fq,\fq)_\infty$.
\begin{figure}[]
\centering
\includegraphics[scale=0.6]{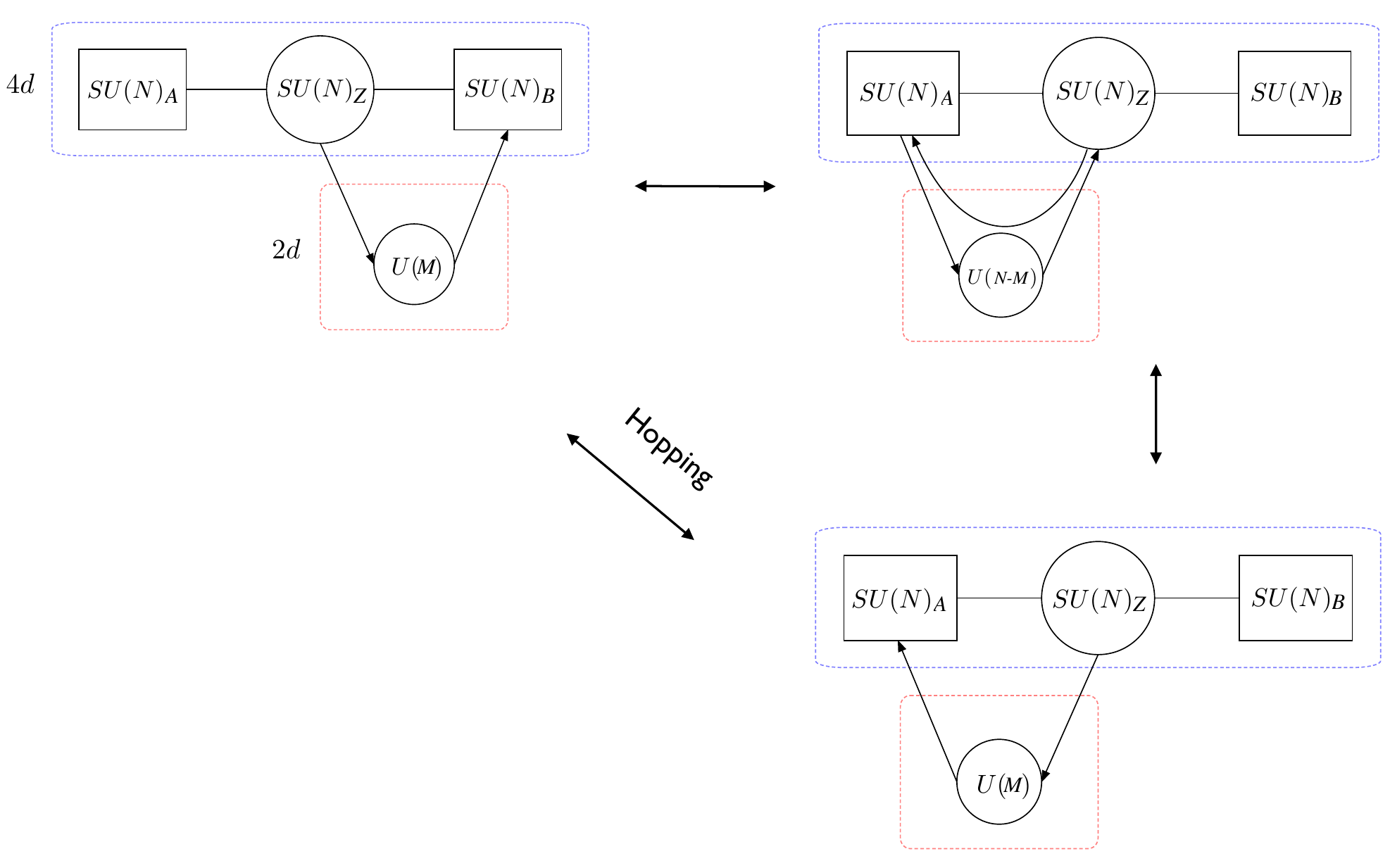}
\caption{An $\cN=(0,2)$ generalization of the ``triality'' considered in \cite{Gadde:2013dda}}
\label{triality}
\end{figure}
After some efforts, it will be proved interesting for us to choose $\fp c^2={\fq^{-(h_{\rm f}+h_{\rm af})}
(\fp\fq)^{\frac{\tr_A\(h_{\rm f}+h_{\rm af}\)+\tfr_{{\rm f}_b}}{2}}}$. 
As a result, the first term in the second line can be absorbed into elliptic gamma function:
\begin{eqnarray}
\cI_{2d-4d}
&=&
\frac{\kappa^{N-1}}{N!}
\oint_{\bbT^{N-1}}
\pl^{N-1}_{\a=1}
\frac{dz_\a}{2\pi i z_\a}
\frac{\pl_{\a,\b=1}^N\Gamma\((\fp\fq)^{\frac{\tr_A}{2}}\frac{z_\a}{z_\b}\)}
{\pl_{\a\neq\b}^N\Gamma\(\frac{z_\a}{z_\b}\)}
\pl_{\a,\b=1}^N
\Gamma\(\frac{z_\a}{\fa_\b}
\(\fp\fq\)^{\frac{\tfr_{{\rm f}_a}}{2}}\)
\Gamma\(\frac{\fa_\b}{z_\a}
\(\fp\fq\)^{\frac{\tfr_{{\rm af}_a}}{2}}\)
\nn\\
&\times&
\sum_{\{\a_l\}}
\pl^N_{\b=1}
\[
\pl_{s\in\{\a_l\}}
\Gamma\(
\frac{z_s}{\fb_\b\fp}
\(\fp\fq\)^{\frac{\tfr_b}{2}}\)
\Gamma\(
\frac{\fb_\b\fp}{z_s}
\(\fp\fq\)^{\frac{\tfr_{{\rm af}_b}}{2}}\)
\pl_{r\in\overline{\{\a_l\}}}
\Gamma\(\frac{z_r}{\fb_\b}
\(\fp\fq\)^{\frac{\tfr_{{\rm f}_b}}{2}}\)
\Gamma\(\frac{\fb_\b}{z_r}
\(\fp\fq\)^{\frac{\tfr_{{\rm af}_b}}{2}}\)
\]
\nn\\
&\times&
\pl_{s\in \{\a_l\}}
\pl_{r\in\overline{\{\a_l\}}}
\frac{\theta(\frac{z_r}{z_s}(\fp\fq)^{\frac{\tr_A}{2}};\fq)}{\theta(\frac{z_r}{z_s};\fq)}.
\end{eqnarray}
After changing the dummy variable  $\tz_s\equiv \fp^{-1}z_s$ and $\tz_r\equiv z_r$, we can readily write down the final expression:
\begin{eqnarray}
\label{2d4d-magnetic}
\cI_{2d-4d}
&=&
\frac{\kappa^{N-1}}{N!}
\oint_{\bbT^{N-1}}
\pl^{N-1}_{\a=1}
\frac{dz_\a}{2\pi i z_\a}
\frac{
\pl_{\a,\b=1}^N\Gamma\((\fp\fq)^{\frac{\tr_A}{2}}\frac{\tz_\a}{\tz_\b}\)}
{
\pl_{\a,\b=1}^N\Gamma\(\frac{\tz_\a}{\tz_\b}\)}
\pl_{\a,\b=1}^N
\Gamma\(\frac{\tz_\a}{\fa_\b}
\(\fp\fq\)^{\frac{\tfr_{{\rm f}_a}}{2}}\)
\Gamma\(\frac{\fa_\b}{\tz_\a}
\(\fp\fq\)^{\frac{\tfr_{{\rm af}_a}}{2}}\)
\Gamma\(\frac{\tz_\a}{\fb_\b}
\(\fp\fq\)^{\frac{\tfr_{{\rm f}_b}}{2}}\)
\Gamma\(\frac{\fb_\b}{\tz_\a}
\(\fp\fq\)^{\frac{\tfr_{{\rm af}_b}}{2}}\)
\nn\\
&\times&
\pl_{\a,\b=1}^N
\frac{\theta\(\frac{\fa_\b}{\fp\tz_\a}
\(\fp\fq\)^{\frac{1-\tfr_{{\rm f}_a}}{2}};\fq\)}
{\theta\(\frac{\fa_\b}{\fp\tz_\a}
\(\fp\fq\)^{\frac{\tfr_{{\rm af}_a}}{2}};\fq\)}
\sum_{\{\a_l\}}
\pl^N_{r,\b=1}
\frac{\theta\(\frac{\tz_r}{\fa_\b}
\(\fp\fq\)^{\frac{1-\tfr_{{\rm af}_a}}{2}};\fq\)}
{\theta\(\frac{\tz_r}{\fa_\b}
\(\fp\fq\)^{\frac{\tfr_{{\rm f}_a}}{2}};\fq\)}
\pl_{s\in \{\a_l\}}
\pl_{r\in\overline{\{\a_l\}}}
\frac{\theta(\frac{\tz_s}{\tz_r}(\fp\fq)^{\frac{\tr_A}{2}};\fq)}{\theta(\frac{\tz_s}{\tz_r};\fq)}.
\end{eqnarray}
Comparing with \eqref{(0,2) Hori-Tong dual}, \eqref{2d4d-magnetic} represents the same 4d theory as \eqref{2d4d-electric} but now coupled with a different 2d degree of freedom which is actually the magnetic theory  for Hori-Tong duality with the flavor fugacity identification shown in the argument:
\begin{eqnarray}
\cI_{2d}&=&
\(\pl_{\a,\b=1}^N
\frac{\theta\(\frac{\fa_\b}{\fp\tz_\a}
\(\fp\fq\)^{\frac{1-\tfr_{{\rm f}_a}}{2}};\fq\)}
{\theta\(\frac{\fa_\b}{\fp\tz_\a}
\(\fp\fq\)^{\frac{\tfr_{{\rm af}_a}}{2}};\fq\)}\)
\cI_{N-M}\(\ta=\tz,\tb=\fa,\tc=\frac{(\fp\fq)^{\frac{\tfr_{{\rm f}_a}+\tr_A\(h_{\rm f}+h_{\rm af}\)}{4}}}{\fq^{\frac{h_f+h_{af}}{2}}};\fq\)\nn\\
&=&\cI_{\rm Magnetic}^{\cN=(0,2)}
\(\ta=\tz,\tb=\fa,
\tc=\frac{(\fp\fq)^{\frac{\tfr_{{\rm f}_a}+\tr_A\(h_{\rm f}+h_{\rm af}\)}{4}}}{\fq^{\frac{h_f+h_{af}}{2}}};\fq\).
\end{eqnarray}
However, from the Hori-Tong duality \eqref{(0,2) Hori-Tong dual}, the coupled 2d magnetic theory also dual to a electric theory whose elliptic genus given as follows:
\begin{equation}
\cI'^{\cN=(0,2)}_{\rm Electric}=\cI_M\(a=\frac{1}{\tz},b=\frac{1}{\fa},
c=\frac{\fq^{\frac{1-(h_{\rm f}+h_{\rm af})}{2}}}{(\fp\fq)^{\frac{\tfr_{{\rm f}_a}+\tr_A\(1-(h_{\rm f}+h_{\rm af})\)}{4}}}\).
\end{equation}
We conclude that the three equivalent 2d-4d coupled theories present above can be regarded as a $\cN=(0,2)$ generalization of the $\cN=(2,2)$ ``triality'' connected by Hori-Tong duality and "Hopping invariance" considered in \cite{Gadde:2013dda}, as illustrated in Figure \ref{triality}.

\acknowledgments
This work was supported in part by National Science Council through the grant No.101-2112-M-002-027-MY3 and Center for Theoretical Sciences at National Taiwan University. Heng-Yu Chen also acknowledges the stimulating discussions with Richard Eager and Kazuo Hosomchi and the hospitality of Kavli Institute for the Physics and the Mathematics of the Universe (KIPMU), University of Tokyo, where parts of this work were completed. The authors would also like to dedicate this work to Sunflower Student Movement in Taiwan, which took place March-April 2014, where many ideas and discussions were inspired while we were sitting among the crowds.

\appendix
\section{Identities for Elliptic Gamma Function and theta function}\label{Appendix-Id}
Here we review some useful identities for the elliptic gamma function and theta function for the use in the main text. {They are defined as}:
\begin{equation}
\label{def-elliptic}
\Gamma(x; p, q) = 
\prod_{r,s \ge 0}
\frac{1-x^{-1} p^{r+1}q^{s+1}}{1-x p^r q^s},
\end{equation}
\begin{equation}\label{def-theta}
\theta(z;p)=\pl^\infty_{l=0}
(1-zp^l)
(1-z^{-1}p^{l+1}).
\end{equation}
We can also define the following ratio between theta functions as 
\begin{equation}\label{def-Delta}
\Delta(x; q, t) = \frac{\theta(xt; q)}{\theta(x; q)}.
\end{equation}
The gamma function inherits an inverse property from its ratio form:
\begin{equation} \label{Inversion-Id}
\Gamma((pq)^{\tilde{r}/2} x; p,q)=\Gamma^{-1}(x^{-1}(pq)^{1-\tilde{r}/2}; p, q) ,
\end{equation}
while similar property for theta function can be found in \cite{Gadde:2013dda}.
These elliptic gamma and theta functions share an elegant relation, obtained by inserting a $p$ (or $q$) into the gamma function argument:
\begin{equation}\label{Gamma-Id1}
\Gamma(pz;p,q)=\theta(z;q)\Gamma(z;p,q)
\end{equation}
which can be generalized into for $m, n \ge 0$:
\begin{eqnarray}
\frac{\Gamma\(xp^nq^m;p,q\)}{\Gamma\(x;p,q\)}
&=&
\frac{1}{(-x)^{mn}}
\frac{1}{p^{\frac{mn(n-1)}{2}}q^{\frac{mn(m-1)}{2}}}
\times
\pl^{n-1}_{r=0}\theta\(xp^r;q\)
\pl^{m-1}_{s=0}\theta\(xq^s;p\).
\label{Gamma-Id4}
\end{eqnarray}
while the same method can be applied to the case shifted by negative integer.
As widely used in the main text, we also write down the residues for elliptic gamma function at $x=p^{-m} q^{-n}$, $m, n=0,1, 2,\dots$:
\begin{equation}\label{Gamma-Res} 
{\rm Res}_{x=p^{-m}q^{-n}}
\left[\Gamma(x; p, q)\right] =
\frac{(-1)^{mn} p^{\frac{nm(m+1)}{2}}q^{\frac{mn(n+1)}{2}}}
{(p;p)_{\infty}(q;q)_{\infty}} 
\prod_{r=1}^{m} \frac{1}{\theta(p^{-r}; q)} \prod_{s=1}^{n} \frac{1}{\theta(q^{-s}; p)}.
\end{equation}
\section{Factorization and Rewriting Integrand into Twisted Superpotential}\label{Appendix-Factorization}
\subsection{Factorization of Elliptic Gamma Function}
\paragraph{}
Here we will use various identities for elliptic gamma function and double sine function given in \cite{Spiridonov-Id} 
to rewrite the integrands in the 4d superconformal index into almost factorizable form.
In appropriate limit the factorizable product reduces to the twisted superpotential obtained from compactifying 4d $\cN=2$ supersymmetric gauge theory on $T^2\times R^2$ \cite{Nekrasov:2009uh}.
Let us first recall from \cite{Aharony:2013dha} that due to the background flux for the global symmetries, the metric on $S^1 \times S^3$ is actually deformed into $\tilde{S}^1\times S^3_b$,
i.e. a $b$-dependent rescaled $\tilde{S}^1$ tof radius $\tilde{r}_1$, trivially fibered over a three dimensional ellipsoid $S_b^3$.
We can also relate the 4d fugacity parameters $\fp, \fq$ and  $z$ with the 3d deformation parameter $b$, 3d radius $r_3$, and the radius of $\tilde{S}^1$ $\tilde{r}_1$ as:
\begin{eqnarray}
&&\fp=e^{2\pi i \tr_1 \w_1};\quad \fq=e^{2\pi i \tr_1 \w_2};\quad z=e^{2\pi i \tr_1 u}\label{3d4dconversion1}\\
&&\tr_1 =\frac{2}{b+\frac{1}{b}}r_1 ;\quad \w_1=ibr_3^{-1};\quad \w_2=ib^{-1}r^{-1}_3 \label{3d4dconversion2}
\end{eqnarray}
as $|\fp|, |\fq| <1$ and $|z|=1$, {where we take $b$ to be real and positive.}
{With the definition given in \cite{Spiridonov-Id} and above, the elliptic gamma function can now be written as:}
\begin{eqnarray}\label{Redef-EllipticGamma}
\Gamma (e^{2\pi i\tr_1 u};e^{2\pi i\tr_1 \w_1},e^{2\pi i\tr_1 \w_2})=
\frac{
\exp\left[\frac{-i\pi}{3}B_{33}(\tr_1 u| 1,\tr_1 \w_1,\tr_1\w_2)\right]
\Gamma_3(\tr_1 u| 1,\tr_1 \w_1,\tr_1\w_2)
        \Gamma_3(1-\tr_1 u| 1,-\tr_1 \w_1, -\tr_1\w_2)}
{\Gamma_3\(1+\tr_1 (\w_1+\w_2-u)| 1,\tr_1 \w_1,\tr_1\w_2\)
\Gamma_3(\tr_1 (u-\w_1-\w_2)| 1,-\tr_1 \w_1, -\tr_1\w_2)},\nn\\
\end{eqnarray}
where various functions are given by:
\begin{equation}
\Gamma_3(x|\Omega_1,\Omega_2,\Omega_3)
=\pl^\infty_{n_1,n_2,n_3=0} (x
+n_1\Omega_1
+n_2\Omega_2
+n_3\Omega_3)^{-1},
\end{equation}
\begin{eqnarray}\label{DefB33-1}
B_{33}(x|\Omega_1,\Omega_2,\Omega_3)&=&
\frac{x^3}{\Omega_1\Omega_2\Omega_3}-
\frac{3(\Omega_1+\Omega_2+\Omega_3)}{2\Omega_1\Omega_2\Omega_3}x^2+
\frac{\sum_{i=1}^3\Omega_i^2
+3(\Omega_1\Omega_2
+\Omega_1\Omega_3
+\Omega_2\Omega_3)}{2\Omega_1\Omega_2\Omega_3}x\nn\\
&&-\frac{1}{4}
\frac{(\Omega_1+\Omega_2+\Omega_3)
(\Omega_1\Omega_2
+\Omega_1\Omega_3
+\Omega_2\Omega_3)
}{\Omega_1\Omega_2\Omega_3}.
\end{eqnarray}
Substituting back into the definition (\ref{Redef-EllipticGamma}) and re-arranging, we obtain that:
\begin{eqnarray}
&&\Gamma(e^{2\pi\tr_1 u}; e^{2\pi i\tr_1 \w_1},e^{2\pi i\tr_1 \w_2})
=\exp\left[\frac{-i\pi}{3}B_{33}(\tr_1 u| 1,\tr_1 \w_1,\tr_1\w_2)\right]\nn\\
&&\times
\(\pl^\infty_{l=0}\pl^\infty_{m,n=0}
\frac{
(1+l-\tr_1 u+(m+1)\tr_1\w_1+(n+1)\tr_1\w_2)\times
(l+\tr_1 u -(1+m)\tr_1\w_1 -(1+n)\tr_1\w_2)}
{
(l+\tr_1 u +m\tr_1\w_1+ n\tr_1\w_2)\times
(1+l-\tr_1 u -m\tr_1\w_1- n\tr_1\w_2)
}\).\nn\\
\end{eqnarray}
Now we can regard the infinite product over "$l$" as coming from the KK mode running around the additional $\tilde{S}^1$ fiber. For each fixed $l$, there actually exists a copy of double-sine function, 
and "$l$" itself can be read as mass of KK mode. Excluding the overall phase factor, we can see this by multiplying all the factors above by $-i(r_3/\tr_1)$ to obtain:
\begin{eqnarray}
\pl^\infty_{m,n=0}
\frac{
\(-i\frac{1+l}{\tr_1}r_3+iur_3+(m+1)b+(n+1)b^{-1}\)
\(i\frac{l}{\tr_1}r_3+iur_3+(m+1)b+(n+1)b^{-1}\)
}
{
\(-i\frac{l}{\tr_1}r_3-iur_3+mb+nb^{-1}\)
\(i\frac{1+l}{\tr_1}r_3-iur_3+mb+nb^{-1}\)
}.\nn\\
\end{eqnarray}
Comparing with the double-sine definition:
\begin{equation}
s_b(c_b+z)=
\pl_{m,n=0}^\infty
\frac{(m+1)b+(n+1)b^{-1}-iz}
       {mb+nb^{-1}+iz},\quad c_b = \frac{i}{2}\(b+\frac{1}{b}\)
\end{equation}
we have:
\begin{eqnarray}
\pl^\infty_{l=0}
s_b\(c_b+\frac{1+l}{\tr_1}r_3-ur_3\)
s_b\(c_b-\frac{l}{\tr_1}r_3-ur_3\)
=\pl_{l \in {\mathbb Z}}s_b\(c_b+\frac{l}{\tr_1}r_3-ur_3\).
\end{eqnarray}
As a result, the elliptic Gamma function can now be written as infinite product of double sine functions:
\begin{equation}\label{FactEGamma}
\Gamma (e^{2\pi i\tr_1 u};e^{2\pi i\tr_1 \w_1},e^{2\pi i\tr_1 \w_2})=\exp\left[\frac{-i\pi}{3}B_{33}(\tr_1 u| 1,\tr_1 \w_1,\tr_1\w_2)\right]
\pl_{l\in{\mathbb Z}}s_b\(c_b+\frac{l}{\tr_1}r_3-ur_3\).
\end{equation} 
As both vector and chiral/hypermultiplet contributions to the 4d superconformal indices can all be expressed in terms of elliptic gamma function $\Gamma(x; \fp, \fq)$, 
while their 3d counterparts are expressed in terms of double sine function $s_b(x)$ \cite{Hama:2011ea}, up to an overall phase, we can therefore regard 4d superconformal index as constructed from summing over infinite number of 3d fields on the ellipsoid $S_b^3$, each labeled by KK mode number $l \in {\mathbb Z}$.
\paragraph{}
On the other hand, we here recall that double-sine function $s_b(x)$ can be further decomposed into two copies of ``holomorphic blocks'' \cite{Pasquetti:2011fj, Beem:2012mb} which are partition function defined on 
$S^1 \times_{\rq} D^2$ where $\rq=e^{2\pi i b^2}$ is the complex structure modulus of the boundary two torus.
The two holomorphic blocks are then glued together via S-fusion rule, $S: {\mathrm q}=e^{2\pi i b^2} \to \tilde{\rq}=e^{2\pi i /b^{2}}$ as:
\begin{eqnarray}
\label{double-sine factorizing}
s_b(c_b-z)=
e^{\frac{-i\pi}{2}(z-c_b)^2}
\pl^\infty_{r=0}
\frac{
1+e^{2\pi i b^2(r+\frac{1}{2})}e^{2\pi b (c_b-z)}}{1+e^{2\pi i b^{-2}(-r-\frac{1}{2})}e^{2\pi b^{-1} (c_b-z)}}
=
e^{\frac{-i\pi}{2}(z-c_b)^2}
||(e^{-2\pi bz}\rq;\rq)_{\infty}||_S^2,\nn\\
\end{eqnarray}
where $(x; q)_{\infty} = \prod_{r=0}^{\infty} (1-x q^r)$ is the Pochhammer symbol.
So again double sine functions $s_b(\pm c_b - z)$ factorize into purely $\rq$ and $\tilde{\rq}$ dependent parts up to an overall phase, for which 
one can further shows that this phase can also be factorized by the following identity for general case:
\begin{equation}\label{S-fusion}
i^\#C^\#
\exp
\[-\frac{1}{2\log \rq}
\((a\cdot X)^2+(i\pi+\frac{\log \rq}{2})b(a\cdot X)\)
\]=
\left|\left|
\theta((-\rq^{\frac{1}{2}})^b)x^a;\rq)_\infty
\right|\right|_S^2,\quad
x=\exp(X)
\end{equation}
where $C$ and $\#$ denote some constant and integer which depends on different ways of factorization and can be absorbed by $||(q;q)_\infty||^2_S=-\frac{2\pi}{\log\rq}C^2$.
For more details and subtleties about factorization, we refer interested readers to \cite{Pasquetti:2011fj, Beem:2012mb, Chen:2013pha}. 
On the other hand, let us now focus on the overall phase factor in (\ref{FactEGamma}). By explicitly using the relations in (\ref{3d4dconversion1}) and (\ref{3d4dconversion2}),
we can factorize $B_{33}$ into purely $b$ and $1/b$ dependence, with $\e\equiv \tr_1 /r_3$ and $\tr_1 u \equiv \tu$:
\begin{eqnarray}\label{DefB33-2}
B_{33}(\tu |1,ib\e,ib^{-1}\e)&=&
-\e^{-2}
\left[\tu^3-\frac{3}{2}\tu^2+\frac{1}{2}\tu-\frac{\e^2}{2}(1+\tu)\right]\nn\\
&&-\e^{-2}
\left[ -\frac{3}{2}i\e b \tu^2
+\frac{3}{2}(i\e b-\e^2b^2)\tu 
-\frac{1}{4}\(i\e b-2\e^2b^2-i\e^3(b^3+b)\)
\right]\nn\\
&&-\e^{-2}
\left[ -\frac{3}{2}i\e b^{-1} \tu^2
+\frac{3}{2}(i\e b^{-1}-\e^2b^{-2})\tu 
-\frac{1}{4}\(i\e b^{-1}-2\e^2b^{-2}-i\e^3(b^{-3}+b^{-1})\)
\right].\nn\\
\end{eqnarray}
We conclude that the elliptic Gamma function is factorizable into purely $b$ and $1/b$ dependent parts, and each part also contains infinite number of KK modes along $\tilde{S}^1$, 
where we can regard the total factorized products as two copies of partition function for a 4d field defined on $D^2 \times T^2$ glued together via S-fusion.

\subsection{Rewriting $\cN=2$ Superconformal Index into twisted superpotential}\label{Appendix:Factorization2}
\paragraph{}
After general discussion in previous section, we perform the concept concretely by considering $\cN=2$ SQCD as the case we did in section \ref{4d-Indices}:
\begin{equation} 
\cI_{\rm 4D}^{\cN=2} = \frac{\ka_4^{N_c}}{N_c!} \oint_{\bbT^{N_c}} \prod_{\a=1}^{N_c}\frac{dz_\a}{2\pi i z_\a} \frac{\prod_{\a,\b=1}^{N_c}\Gamma\left(\frac{z_\a}{z_\b}\frac{\fp\fq}{\ft}; \fp, \fq\right)}{\prod_{\a\neq \b}^{N_c} \Gamma(\frac{z_\a}{z_\b}; \fp, \fq)} \prod_{i=1}^{N_f} \prod_{\a=1}^{N_c}
\frac{
\Gamma\left(\frac{z_\a }{\fa_i}\sqrt{\ft}; \fp, \fq\right) }
{\Gamma\left(\frac{z_\a}{\fa_i}\frac{\fp\fq}{\sqrt{\ft}}; \fp, \fq\right) }.
\end{equation} 
Now substituting the redefinition of parameters in (\ref{FactorizationParameters})
and applying\eqref{FactEGamma}, we can rewrite the contour integral into following alternative form:
\begin{eqnarray} 
\cI_{\rm 4D}^{\cN=2} &=& \frac{\ka_4^{N_c}}{N_c!} \oint_{\bbT^{N_c}} \prod_{\a=1}^{N_c}\frac{dz_\a}{2\pi i z_\a} 
\prod_{\a,\b=1}^{N_c}
\[
e^{\frac{-i\pi}{3}\[B_{33}\(\s_\a-\s_\b+2\e c_b-i\e J)\)-
B_{33}\(\s_\a-\s_\b\)
\]}
\pl^{\infty}_{l=-\infty}
\frac{s_b\(-c_b+2iJ+\e^{-1}(l-\s_\a+\s_\b)\)}
      { s_b\(c_b+\e^{-1}(l-\s_\a+\s_\b)\)}\]
\nn\\
&&\qquad\times
\prod_{i=1}^{N_f} \prod_{\a=1}^{N_c}
\[
e^{\frac{-i\pi}{3}\[B_{33}\(\sigma_\a-m_i+i\e J \)-B_{33}\(\sigma_\a-m_i+2\e c_b-i\e J \)\]}
\pl^\infty_{l=-\infty}
\frac{
s_b\(
c_b-iJ
+\e^{-1}\(l-\sigma_\a+m_i\)
\)}
{
s_b\(
-c_b+iJ
+\e^{-1}\(l-\sigma_\a+m_i\)\)}\].\nn\\
\end{eqnarray} 
Let us pause here to discuss what we want to identify as the "holomorphic parameters" and the "anti-holomorphic parameters". Recall that we have two complementary ways to insert a two torus $T^2$ into four dimensions corresponds to two different parameter matching conditions \eqref{matchCond1} and \eqref{matchCond2}, which give the matching of 2d and 4d fugacity parameters \eqref{N=2Matching} and its counterpart under $\fp \leftrightarrow \fq$ exchange:
\begin{equation}
p=\fp, \quad t=\frac{\fp\fq}{\ft}, \quad d'=\frac{\fq^2}{\ft} .
\end{equation}
These inspire us to consider the following two sets of combinations of parameters in the holomorphic copy and anti-holomorphic copy respectively, which are exchanged under $b\leftrightarrow b^{-1}$:
\begin{eqnarray}
{\rm Holomorphic :} (ib, c_b-i J, i b^{-1}-i J),\label{Block parameter1}\\
{\rm Anti-holomorphic :} (ib^{-1}, c_b-iJ, ib-iJ).
\label{Block parameter2}
\end{eqnarray}
For simplicity, in the following, we define $\D=c_b-i J$ and factorize the integrand into blocks characterized by above two categories of parameters. We focus on the double-sine function part first. Contribution from the vector multiplet, accompanied by \eqref{double-sine factorizing} and \eqref{S-fusion}, can be regulated into S-fusion of two holomorphic blocks. Furthermore, to rewrite them into the twisted superpotential form, we use the identity:
\begin{equation}
(\rq z;\rq)_\infty
\rightarrow
\exp\(\frac{1}{\log \rq}\sum^\infty_{m=0}\frac{B_m(\log \rq)^m}{m!}{\rm Li_{2-m}(z)}\) \quad{\rm as}\quad \rq\rightarrow 1^- .
\end{equation}
Skipping the calculation details, we have the vector multiplet contributions:
\begin{eqnarray}\label{FactVec}
&&\sum_{\a,\b=1}^{N_c}
\sum_{l\in {\mathbb Z}}
\log\left[\frac{s_b\(-c_b+2i J-\s_\a+\s_\b+\e^{-1}l\)}
      { s_b\(c_b-\s_\a+\s_\b+\e^{-1}l\)}\right]
\nn\\
&&=
\sum_{\a,\b=1}^{N_c}\sum_{l\in {\mathbb Z}}\sum_{m=0}^{\infty}
\frac{1}{\log \rq}\frac{B_m(\log \rq)^m}{m!}
\[
-
{\rm Li_{2-m}}\(e^{2\sqrt{2}\pi b {\D}}\)
+
{\rm Li_{2-m}}\(e^{2\pi b\(-\s_\a+\s_\b+ \frac{l}{\e}\)-2\D}\rq\)
-
{\rm Li_{2-m}}\(e^{2\pi b \(-\s_\a+\s_\b+\frac{l}{\e}\)}\rq\)
\]\nn\\
&&+ (b\rightarrow b^{-1}),
\end{eqnarray}
and the same goes for the contributions from hypermultiplets:
\begin{eqnarray}\label{FactHyper}
&&\sum_{\a=1}^{N_c}
\sum_{i=1}^{N_F}
\sum_{l\in {\mathbb Z}}
\log\left[\frac{
s_b\(
c_b-i J
-\sigma_\a+m_i
+\frac{l}{\e}
\)}
{
s_b\(
-c_b+iJ
-\sigma_\a+m_i
+\frac{l}{\e}\)}\right]\nn\\
&&=
\sum_{\a=1}^{N_c}
\sum_{i=1}^{N_F}
\sum_{l\in{\mathbb Z}}
\sum_{m=0}^{\infty}
\frac{1}{\log \rq}
\frac{B_m(\log \rq)^m}{m!}
\[
{\rm Li_{2-m}}\(e^{2\pi b
(-\sigma_\a+m_i+\D)}\rq\)
-
{\rm Li_{2-m}}\(e^{2\pi b
(-\sigma_\a+m_i)}\rq\)
-
{\rm Li_{2-m}}\(e^{2\pi b
{\D}}\rq\)
\right.
\nn\\
&&\qquad\qquad\qquad\qquad\qquad\qquad\qquad
\left.
+
{\rm Li_{2-m}}\(e^{2\pi b\(
-\s_\a+m_i+i(b^{-1}-J)+
\frac{l}{\e}\)}\rq\)
-
{\rm Li_{2-m}}\(e^{2\pi b\(
-\s_\a+m_i+i(b^{-1}-J)+
\frac{l}{\e}-2\D\)}\rq\)
\]\nn\\
&&
+ (b\rightarrow b^{-1}).
\end{eqnarray}
It is interesting to note that in (\ref{FactVec}) and (\ref{FactHyper}), we can further split into $l$ independent and dependent pieces, representing purely 3d degrees of freedom and KK modes along the $\tilde{S}^1$. 
However, the factorization property above does not manifest when we consider the term consists of $B_{33}$ function which encodes the additional degree of freedom when ellipsoid $S^3_b$ is trivially fibered by $\tilde{S}^1$. To make it more explicit, we shall follow the dictionary \eqref{Block parameter1} and \eqref{Block parameter2} and write down following expressions: 
\begin{eqnarray}
&\pl_{\a,\b=1}^{N_c}&
\exp\(\left.\left.\frac{-i\pi}{3}
\right[ B_{33}\(\e\[\sigma_\a-\s_\b+2c_b -2i J\] \)-B_{33}\(\e\[\sigma_\a-\s_\b\] \)
\]\)\nn\\
=
&\pl_{\a,\b=1}^{N_c}&
\exp\(
-i\pi\D
\[\frac{8}{3}\e\D^2+2\e(\sigma_\a-\s_\b)^2
-2(1-2\e c_b)\D+(\frac{\e^{-1}-b^2\e-b^{-2}\e}{3}+(2 c_b-\e))
\]\right)
\nn\\
=
&\pl_{\a,\b=1}^{N_c}&
\exp\(
-i\pi\D
\[\frac{8}{3}\e\D^2+2\e(\sigma_\a-\s_\b)^2
-2\D+\frac{\e^{-1}}{3}-\e
+(\frac{-b^2\e}{3}
+2i\e b\D
+ib)
+(\frac{-b^{-2}\e}{3}
+2i\e b^{-1}\D
+ib^{-1})
\]\right)
\nn\\
\end{eqnarray}
and
\begin{eqnarray}
&\pl_{\a=1}^{N_c}\pl^{N_F}_{i=1}&
\exp\(\frac{-i\pi}{3}
\[B_{33}\(\e\[\sigma_\a-m_i+i J\] \)-B_{33}\(\e\[\sigma_\a-m_i+2c_b-i J\] \)
\]\)\nn\\
=
&\pl_{\a=1}^{N_c}\pl^{N_F}_{i=1}&
\exp\(
{-i\pi\D}
\[\frac{8\e}{3}\D^2
+4\e\D (\sigma_\a-m_i)
+2\e(\sigma_\a-m_i)^2
-2\D
-2(\sigma_\a-m_i)+
\frac{\e^{-1}}{3}-\e
\right.\right.\nn\\
&&\qquad\qquad\left.
+\e(J-ib)^2+(2\e\D+2\e(\sigma_\a-m_i)-1)(J-i b)+\frac{2}{3}\e b^{-2}\right.\nn\\
&&\qquad\qquad\left.\left.
+\e(J-i b^{-1})^2+(2\e\D+2\e(\sigma_\a-m_i)-1)(J-i b^{-1})+\frac{2}{3}\e b^{2}\]\).
\end{eqnarray}
By this way, we can still identify part of them as product of holomorphic and anti-holomorphic block while the non-factorizable terms are extracted now. Note that they have a common dependence on 
$\D=\frac{1}{4i\pi \e}\log\frac{\fp\fq}{\ft}$, this reflects the fact that in \eqref{4DN=2Index-Int} the non-factorizable term comes out as power of $\frac{\fp\fq}{\ft}$.   Without further discussions for the time being, we treat it simply as a "gluing factor" to glue up two copies of $D^2\times T^2$ under S-fusion. 
Combine all factors and we can get the factorized twisted superpotential form:
\begin{eqnarray}
&&{\cI}^{\cN=2}=\frac{\kappa_4^{N_c}}{N_c !} \int\pl^{N_c}_{\a=1}d\s_\a
\pl^{N_c}_{\a,\b=1}
\exp\(
-i\pi\D
\[\frac{8}{3}\e\D^2+2\e(\sigma_\a-\s_\b)^2
-2\D+\frac{\e^{-1}}{3}-\e\]\)\nn\\
&&\qquad\qquad
\times
\pl_{\a=1}^{N_c}\pl^{N_F}_{i=1}
\exp\(
{-i\pi\D}
\[\frac{8\e}{3}\D^2
+4\e\D (\sigma_\a-m_i)
+2\e(\sigma_\a-m_i)^2
-2\D
-2(\sigma_\a-m_i)+
\frac{1}{3\e}-\e\]\)
\nn\\
&&
\times
\left\{
\pl^{N_c}_{\a,\b=1}
\exp\(-i\pi \D
\[\frac{-b^2\e}{3}
+2i\e b\D
+ib\]\)\right.\nn\\
&&
\times
\pl^{N_c}_{\a=1}
\pl^{N_F}_{i=1}
\exp\(-i\pi\D
\[\e(J-ib^{-1})^2+(2\e\D+2\e(\sigma_\a-m_i)-1)(J-i b^{-1})+\frac{2}{3}\e b^2\]\)
\nn\\
&&
\times
\pl^{N_c}_{\a,\b=1}
\pl_{l\in {\mathbb Z}}
\exp\(\sum^\infty_{m=0}\frac{B_m(\log \rq)^{m-1}}{m!}
\[
-
{\rm Li_{2-m}}\(e^{2\sqrt{2}\pi b {\D}}\)
+
{\rm Li_{2-m}}\(e^{2\pi b \(\s_\b-\s_\a+\frac{l}{\e}-2\D\)}\rq\)
-
{\rm Li_{2-m}}\(e^{2\pi b\( \s_\b-\s_\a+\frac{l}{\e}\)}\rq\)
\]\)\nn\\
&&
\times
\pl^{N_f}_{i=1}
\pl^{N_c}_{\a=1}
\pl_{l\in{\mathbb Z}}
\exp\(
\sum^\infty_{m=0}
\frac{B_m(\log \rq)^{m-1}}{m!}
\[
{\rm Li_{2-m}}\(e^{2\pi b
(-\s_\a+m_i+\D)}\rq\)
-
{\rm Li_{2-m}}\(e^{2\pi b
(-\s_\a+m_i)}\rq\)
-
{\rm Li_{2-m}}\(e^{2\pi b
{\D}}\rq\)
\right.\right.
\nn\\
&&\quad\qquad\qquad\qquad\qquad\qquad\qquad
\left.\left.
\left.+
{\rm Li_{2-m}}\(e^{2\pi b\(
-\s_\a+m_i+i(b^{-1}-J)+
\frac{l}{\e}\)}\rq\)
-
{\rm Li_{2-m}}\(e^{2\pi b\(
-\s_\a+m_i+i(b^{-1}-J)+
\frac{l}{\e}-2\D\)}\rq\)
\]\)\right\}\nn\\
&&
\times \{b\rightarrow b^{-1}\}.
\end{eqnarray}
For consistency, it can be checked that by taking the limit $\e\rightarrow 0$, we can recover the 3d results on two copies of $D^2\times S^1$ \cite{Chen:2013pha} while KK-modes with non-zero "$l$"  decoupled, and, up to a divergence, the non-factorizable terms reduce to the form of \eqref{S-fusion}, becoming factorizable as we expect.
Moreover we can consider the degenerate limit such that $b \to 0$, while keeping fixed $\frac{b}{\epsilon}$ and $\epsilon J$, along with the $\{b\sigma_\alpha\}$ and $\{b m_i\}$.   
In this case, when we  remove the $1/b$ dependent pieces in the last line above, and only keep $m=0$ term in the summation, 
up to overall phase, we recover the form of twisted superpotential for 4d $\cN=2$ gauge theory with the identical gauge group and matter contents reduced on $T^2 \times R^2$
with $\frac{b}{\epsilon}$ being the complex structure modulus of $T^2$. We can also consider the complementary limit of $b^{-1} \to 0$, and keep other corresponding quantities fixed to recover the other copy of reduction on $T^2\times R^2$ with complex structure of $T^2$ being $\frac{1}{\epsilon b}$.

\end{document}